\documentclass[10pt]{article}
\usepackage{graphicx,psfrag,amsmath,amssymb,amsfonts,bbm,latexsym}
\begin{document}

\title{
Ultracold atomic Fermi-Bose mixtures \\ 
in bichromatic optical dipole traps: \\
a novel route to study fermion superfluidity}

\author{ Roberto Onofrio 
\thanks{Dipartimento di Fisica ``G. Galilei'', Universit\`a di Padova, 
Via Marzolo 8, Padova 35131, Italy} \thanks{Istituto Nazionale per la 
Fisica della Materia, Unit\`a di Roma 1 and Center for Statistical 
Mechanics and Complexity, Roma 00185, Italy} \thanks{Los Alamos
National Laboratory, Los Alamos, NM 87545}
and Carlo Presilla \thanks{Dipartimento di Fisica,
Universit\`a di Roma ``La Sapienza'', Piazzale A. Moro 2, Roma 00185,
Italy} \thanks{Istituto Nazionale per la Fisica della Materia, Unit\`a
di Roma 1 and Center for Statistical Mechanics and Complexity, Roma
00185, Italy} \thanks{Istituto Nazionale di Fisica Nucleare, Sezione
di Roma 1, Roma 00185, Italy} }

\date{\today} \maketitle

\begin{abstract}
The study of low density, ultracold atomic Fermi gases is a promising
avenue to understand fermion superfluidity from first principles.   
One technique currently used to bring Fermi gases in the degenerate 
regime is sympathetic cooling through a reservoir made of an ultracold 
Bose gas. 
We discuss a proposal for trapping and cooling of two-species 
Fermi-Bose mixtures into optical dipole traps made from combinations 
of laser beams having two different wavelengths. 
In these bichromatic traps it is possible, by a proper choice of 
the relative laser powers, to selectively trap the two species 
in such a way that fermions experience a stronger confinement 
than bosons.
As a consequence, a deep Fermi degeneracy can be reached 
having at the same time a softer degenerate regime for the Bose gas.
This leads to an increase in the sympathetic cooling efficiency 
and allows for higher precision thermometry of the Fermi-Bose mixture.
\end{abstract}



\section{Introduction}
Superfluidity and superconductivity are phenomena at the heart of
quantum mechanics of many-body systems.
Their importance is not limited to condensed matter physics 
and some of the concepts involved in their
understanding have been seminal in other contexts, most notably in
quantum field theory \cite{NAMBU}. A wealth of experimental
informations on macroscopic quantum transport has been collected 
by studying the Bose and Fermi isotopes of helium in the superfluid 
phase and the electron liquids in superconductors \cite{TILLEY}. 
With the advent of Bose-Einstein condensates of dilute 
gases consequent to the development of innovative cooling technologies 
\cite{CHU,COHEN,PHILLIPS,KETTERLE}, the class of systems available for 
studying superfluidity has been enlarged to many other species, 
namely various isotopes and hyperfine states of alkalis as well as 
atomic hydrogen, metastable helium, and ytterbium. 
Besides the richer spectrum of investigable atomic systems, 
in the case of dilute gases one can exploit their intrinsically 
slower dynamics - low densities imply weak average interactions - 
to study formation and decay of interesting structures like vortices, 
a possibility very hard to achieve in the denser liquid helium.
Moreover, unlike liquid helium, the precision achievable with atomic 
physics experimental techniques, and the possibility to control
the theoretical many-body approximations, are two other reasons which 
have contributed to the fast development of this sector at 
the borderline between atomic and condensed matter physics.

After the pioneering observation of Bose-Einstein condensation (BEC) 
in dilute gases of $^{87}$Rb \cite{ANDERSON}, 
$^{23}$Na \cite{DAVIS}, and ${}^7$Li \cite{BRADLEY}, 
a lot of experimental and theoretical activities have been focused on 
the signatures of superfluid behaviour in this novel low-density 
state of matter (see for instance \cite{ASPECT,PETHICK,PITAEVSKIIBOOK}). 
Examples are the formation of vortices by means of optomechanical
driving \cite{MATTHEWS} and mechanical stirring \cite{MADISON,ABOSHAEER},
the spectroscopy of scissor modes \cite{MARAGO}, 
the studies of superfluid flow and the related onset of a
critical velocity \cite{RAMAN,ONOFRIO}. 
These experimental achievements have been complemented by relatively 
simple, first-principle theoretical studies with significant 
progress in the comprehension of longstanding issues in liquid $^4$He, 
like critical velocities and vortex formation \cite{FETTER,LEGGETT}. 
Even more interesting is the study, still completely open from 
the experimental viewpoint, of superfluid features in Fermi 
dilute gases \cite{PITAEVSKIISCIENCE}. 
The presence of a superfluid phase is expected to occur in the deep 
degenerate regime via a sort of atomic Cooper pairing
\cite{GORKOV,STOOF}, on the basis of qualitative analogies to the 
case of ${}^3$He and, more in general, to high density electron 
liquids in superconductors. 

In this paper we discuss in some length a recent proposal to 
confine and cool two-species Fermi and Bose gases in a bichromatic 
optical dipole trap \cite{ONPRE}. This configuration allows for selective
trapping of the two species with different trapping strengths. 
Since the confinement determines the degeneracy conditions, 
a regime can be chosen such that the Fermi temperature is 
much larger than the Bose-Einstein condensation temperature. 
In this case, a deep Fermi degeneracy could be achieved before 
(or in proximity) of the BEC phase-transition for the Bose component, 
leading to various advantages in the search for superfluidity 
in dilute fermions amidst a thermal (or thermally-dominated) Bose gas. 
This would provide an unprecedented situation as compared 
to the already available ${}^3$He-${}^4$He mixture. 
In principle, many Fermi-Bose mixtures can be studied. 
We will focus the discussion on the best two Bose coolers available, 
${}^{23}$Na and ${}^{87}$Rb - the workhorses for Bose-Einstein
condensation - and the only two stable Fermi isotopes for alkali
atoms, ${}^6$Li and ${}^{40}$K. 

The paper is organized as follows.
In Section 2 we briefly describe the experimental techniques specific
to the trapping and cooling of Fermi dilute gases, and then give an 
updated overview of the current experimental efforts in reaching Fermi
degeneracy. In Section 3 we introduce our proposal by discussing the
conservative trapping features in various configurations and for
diverse combinations of Fermi-Bose mixtures. In Section 4 we discuss
evaporative cooling for the boson species by giving a specific example
of its dynamics in the case of one-color and two-color single-beam 
optical dipole traps. In Section 5 we deal with sympathetic cooling, 
with particular regard to the limitations to the minimum achievable 
temperatures induced by the heat capacities of the two species.
In the same Section, we discuss qualitatively also some methods to 
evidence a possible superfluid state in the Fermi component, and
comment on the advantages of having a thermal Bose cloud rather than 
the only Bose condensed component.  General features of our proposal 
are summarized and discussed in the conclusions.

\section{Status of the experimental searches for superfluidity in 
dilute Fermi gases} 

As a general consequence of the Heisenberg principle, quantum degeneracy
occurs at temperatures very similar for bosons and fermions in
presence (or absence) of external confining potentials with comparable
strength. It is therefore natural to apply to the Fermi gases the
successful trapping and cooling techniques already developed for
bosons and culminating in the observation of a Bose condensed phase.
However, there are also very striking differences between bosons and
fermions in the degenerate regime. For instance, fermions
enter into a degenerate regime without the sharp phase transition
characteristic of bosons. 
Also, when dealing with Fermi systems one has always to face the 
effects of the Pauli principle which freezes most of
the available degrees of freedom.
The effects of the Pauli principle are particularly felt in all the current
experimental efforts to achieve full degeneracy and to evidence a
superfluid phase in Fermi systems confined by means of magnetic traps. 
With this trapping technique only spin-polarized Fermi gases can be 
confined and cooling is ultimately obtained - like in their bosonic 
counterparts - by the selective removal of the most energetic part of 
the atomic cloud allowing for rethermalization of the remaining fraction. 
Consequently, since the atoms are polarized in the same hyperfine state, 
the Pauli principle forbids the $s$-wave elastic scattering. 
Then rethermalization becomes inefficient at low temperature where the 
contribution of odd angular momentum states (like $p$-wave scattering) 
is strongly suppressed, and this limits the efficiency of direct 
evaporative cooling among fermions in the same hyperfine state.
Two routes to overcome this limitation have been implemented, dual
evaporative cooling of fermions prepared in two different hyperfine 
states, and sympathetic cooling with a Bose refrigerant.   
Even in these situations the Pauli principle gives limitations,  
as the elastic scattering between different hyperfine states is 
inhibited by the so-called Pauli blocking \cite{DEMARCO}:
the atoms available to elastic scattering are limited to the Fermi 
surface and their number is directly proportional to $T/T_\mathrm{F}$. 
The use of sympathetic cooling with a boson reservoir is instead
obstacled by the superfluidity of the latter \cite{TIMMERMANS}, 
the fact that the specific heat of the Bose gas quickly vanishes 
fot $T < T_\mathrm{c}$, and ultimately by Pauli blocking. 
As a matter of fact, the lowest temperatures presently achieved for
fermions are limited in the 0.05-0.2 $T_\mathrm{F}$ range
\cite{TRUSCOTT,SCHRECK,GRANADE,HADZIBABIC,ROATI,GEHM,REGAL,STRECKER,HADZIBABIC1,CUBIZOLLES,GOLDWIN}.

Concerning the trapping techniques, the use of magnetic traps 
gives limitations in the combinatorics of trappable hyperfine states 
and interferes with the use of tunable homogeneous magnetic fields 
required to enhance atomic scattering via Feshbach resonances as 
predicted in \cite{FESHBACH} (see also \cite{TIMMERMANSREV} for 
a recent review) and observed in various atomic systems 
\cite{INOUYE,COURTEILLE,ROBERTS,VULETIC}. 
Feshbach resonances provide a mechanism that could be crucial to identify 
signatures of superfluidity even at relatively large temperatures, the 
so-called {\sl resonant} superfluidity 
\cite{TIMMERMANS1,HOLLAND,CHIOFALO,OHASHI}. 

Some of the above mentioned limitations can be overcome by 
using optical dipole traps \cite{ASKIN,GORDON,CHU1,MILLER}. 
Both different hyperfine states and arbitrary magnetic fields can be 
used in this case. 
Optical dipole traps have been pursued as a way to obtain quantum
degeneracy with purely optical tools, avoiding the complications of 
magnetic trapping \cite{ADAMS}. 
After studies on degenerate Bose gases generated in magnetic traps
and then transferred into optical dipole traps \cite{STAMPER}, 
both BEC \cite{BARRETT,GRIMM} and Fermi degeneracy \cite{GRANADE} 
have been achieved directly in all-optical traps. 
Also, preliminary studies of Feshbach resonances for fermions in an 
optical trap have been reported in \cite{LOFTUS}. 
More recently, studies of strongly interacting Fermi gases have been 
reported in the case of Fermi-Bose mixtures \cite{MODUGNOLAST} 
and two-component Fermi gases \cite{OHARA0,NOTETTF}. 
In the latter case the expansion dynamics has been
interpreted as a possible evidence of resonant superfluidity 
as predicted in \cite{MENOTTISTRING}.

The experimental situation, updated to the Summer 2003, 
is summarized in Table 1. 
\begin{table}[h]
\begin{center}
\begin{tabular}{crcl}
\hline\hline
Atomic species & $N_\mathrm{F}$ & $T/T_\mathrm{F}$ & Ref. \\ 
\hline  
$^{40}$K($9/2,9/2$) - $^{40}$K($9/2,7/2$) & $7 \times 10^5$ & 0.50 & JILA1  \cite{DEMARCO} \\ 
$^{6}$Li($3/2,3/2$) - $^{7}$Li($2,2$) & $1.4 \times 10^5$ & 0.25 & Rice1\cite{TRUSCOTT} \\
$^{6}$Li($3/2,3/2$) - $^{7}$Li($2,2$) & $4 \times 10^3$ & 0.20 & ENS1 \cite{SCHRECK} \\ 
$^{6}$Li($1/2,1/2$) - $^{6}$Li($1/2,-1/2$) & $10^5$ & 0.50 & Duke1 \cite{GRANADE} \\ 
$^{6}$Li($1/2,1/2$) - $^{23}$Na($1,-1$) & $1.4 \times 10^5$ & 0.50 &MIT1\cite{HADZIBABIC} \\ 
$^{40}$K($9/2,9/2$) - $^{87}$Rb($2,2$) &  $10^4$&0.30&LENS\cite{ROATI} \\
$^{6}$Li($1/2,1/2$) - $^{6}$Li($1/2,-1/2$) & $1.6 \times 10^5$ & 0.15 & Duke2 \cite{GEHM} \\ 
$^{40}$K($9/2,-9/2$) - $^{40}$K($9/2,-5/2$) & $1.1 \times 10^6$ & 0.21 & JILA2  \cite{REGAL} \\ 
$^{6}$Li($3/2,3/2$) - $^{7}$Li($2,2$) & $7 \times 10^7$ & 0.10 &Rice2\cite{STRECKER} \\
$^{6}$Li($1/2,1/2$) - $^{23}$Na($2,2$) & $7 \times 10^7$ & 0.05 &MIT2\cite{HADZIBABIC1} \\ 
$^{6}$Li($1/2,-1/2$) - $^{6}$Li($1/2,1/2$) & $8 \times 10^4$ & 0.43 & ENS2 \cite{CUBIZOLLES} \\ 
\hline\hline
\end{tabular}
\end{center}
\caption{Status of the experimental studies of degenerate Fermi dilute
gases. For each laboratory the various trapped species are reported
specifying the particular  hyperfine state $(F,m_F)$, 
the number of fermions at the final stage of cooling $N_\mathrm{F}$, 
the lowest reached temperature ratio $T/T_\mathrm{F}$, 
and the related reference. 
The quoted experiments at JILA and Duke University make use of dual 
evaporative cooling of two hyperfine states of magnetically trapped 
potassium and optically trapped lithium, respectively. 
The other experiments exploit sympathetic cooling with bosonic
reservoirs, $^{7}$Li,$^{23}$Na, and $^{87}$Rb. 
More recently, sympathetic cooling of $^{40}$K 
with $^{87}$Rb has been also pursued at JILA \cite{GOLDWIN}.  
In subsequent work the Duke University group has reached deeper degeneracy and observed
an anisotropic expansion of the Fermi gas, which could be interpreted as 
an evidence of superfluid behavior of the Fermi gas 
\cite{OHARA0,NOTETTF}. More recent experiments use Feshbach
resonances, resulting in high-efficiency production of potassium or lithium
ultracold molecules. In the experiment described in \cite{STRECKER} the Fermi-Bose 
mixture is an intermediate stage, then all the bosons are removed and the 
fermions are prepared in an incoherent mixture of equal populations in the 
$(1/2,-1/2)$ and $(1/2,1/2)$ states.}
\end{table}
The two stable fermionic species $^{6}$Li and $^{40}$K have been cooled 
by using evaporative cooling between two hyperfine states 
(JILA and Duke University) or through sympathetic cooling with Bose 
condensates of $^{7}$Li, $^{23}$Na, and $^{87}$Rb 
(Rice-ENS, MIT, and LENS respectively). 
It is evident that, despite of very different trapping and cooling
techniques, the lowest degeneracy parameter $T/T_\mathrm{F}$ 
obtained by using sympathetic cooling is around 0.2  
(for the particular case of MIT2 see a detailed discussion in Section 5).
The existence of this sort of lower bound for $T/T_\mathrm{F}$ can be
understood semiquantitatively in the following way \cite{PRESILLA}.
In the latest stage of evaporative cooling the number of Bose atoms 
becomes of the same order of magnitude of the number of Fermi atoms 
(in some hyperfine states of $^7$Li, due to the negative scattering
length, there is also a theoretical upper limit to the number of atoms 
in the condensed phase, see \cite{BRADLEY}). 
On the other hand the boson specific heat scales as $T^3$ below the 
BEC transition, while the Fermi atoms have specific heat scaling as $T$. 
The specific heat curves of bosons and fermions, assuming for
simplicity equal masses for the two species, intersect each other
at a temperature $T^* \sim 0.5 ~T_\mathrm{c} \sim 0.25 ~T_\mathrm{F}$. 
Below $T^*$ sympathetic cooling becomes very inefficient.
This simple explanation gives also an hint on how to overcome the limitation. 
If one could engineer the trapping potential in such a way that 
$T_\mathrm{c} \ll T_\mathrm{F}$, the Bose gas would preserve 
enough thermal capacity to drive the sympathetic cooling even 
in a deep degenerate regime for the Fermi component. 
A first attempt in this direction can be found in \cite{VIVERIT}, 
where adiabatic compression was proposed in an optical
dipole trap superimposed to an already confining magnetic trap.  
This should allow a small fraction of the Fermi atoms to experience 
a very tight confinement potential, thus enhancing the Fermi temperature.
More recently, we have proposed an alternative solution in the context 
of pure optical trapping \cite{ONPRE}, by using a two-color optical
dipole trap which confines both Fermi and Bose gases with different 
strengths. It is the purpose of the next Section to discuss in detail 
the static confinement features of this class of atomic traps.

\section{Two-color optical dipole traps}
Let us start our analysis from the conditions of degeneracy for Fermi
and Bose gases confined in harmonic traps. The Fermi and Bose
temperatures for dilute atomic clouds trapped by harmonic potentials
can be written as:
\begin{equation}
T_\mathrm{F}= 6^{1/3}~
\hbar \omega_\mathrm{f} N_\mathrm{f}^{1/3} k_\mathrm{B}^{-1}
\simeq 1.82 ~
\hbar \omega_\mathrm{f} N_\mathrm{f}^{1/3} k_\mathrm{B}^{-1}
\label{TF}
\end{equation}
\begin{equation}
T_\mathrm{c} = \zeta(3)^{-1/3}~
\hbar \omega_\mathrm{b} N_\mathrm{b}^{1/3} k_\mathrm{B}^{-1}
\simeq 0.94 ~
\hbar \omega_\mathrm{b} N_\mathrm{b}^{1/3} k_\mathrm{B}^{-1}
\label{Tc}
\end{equation}
with $\omega_\mathrm{f}=(\omega_{\mathrm{f}x} \omega_{\mathrm{f}y} 
\omega_{\mathrm{f}z})^{1/3}$
and
$\omega_\mathrm{b}=(\omega_{\mathrm{b}x} \omega_{\mathrm{b}y} 
\omega_{\mathrm{b}z})^{1/3}$ being the geometrical average of the 
angular trapping frequencies in the three directions for fermions and 
bosons, $N_\mathrm{f}$ and $N_\mathrm{b}$ the number of atoms of the 
Fermi and Bose gases, and $\hbar$ and $k_\mathrm{B}$ the Planck and 
Boltzmann constants, respectively.

Besides the small difference in the prefactor, the degenerate
temperatures for Fermi and Bose atoms per unit of atom are
similar in traps with the same angular trapping frequencies for the
two species. This is indeed the case of magnetic traps: since
the magnetic moments of the alkali-metals are very similar, the only
difference in the trapping strengths is due to their different
masses, with the angular trapping frequencies scaling as 
$\omega_\mathrm{f}/\omega_\mathrm{b} \simeq 
(m_\mathrm{b}/m_\mathrm{f})^{1/2}$. 
The situation may change in optical dipole
traps where the confinement is dictated by the detunings of
the laser beams with respect to the atomic transitions and 
by the beam intensities. 
Situations for which Fermi degeneracy is reached before BEC 
(\textit{i.e.} $T_\mathrm{F} > T_\mathrm{c}$) are therefore viable, 
provided that fermions and bosons have different atomic transitions. 
This will restrict our analysis to two-species mixtures, since the 
isotopic shifts are usually not enough to ensure selective trapping 
in single species Fermi-Bose mixtures without incurring in
prohibitive heating due to residual Rayleigh scattering. 

In discussing optical dipole traps, one can consider either single
beam or crossed-beam configurations \cite{ADAMS}, see Figure 1. 
The former has the advantage of being simpler with fewer
experimental problems of loading and alignment, the latter gives rise 
to a more isotropic confinement. In the following, we will discuss 
in detail the crossed-beam geometry. Results for the single beam
configuration will be obtained as a particular case in which one of 
the two beams is turned off. 
\begin{figure}
\begin{center}
\vspace{2cm}
\textsf{size of eps file too large, not accepted by arXiv}
\\
\textsf{Curious to see the figure?
look at the published paper 
http://dx.doi.org/10.1023/B:JOSS.0000019829.71660.40}
\vspace{2cm}
\caption{Combined optical dipole trap for two-species mixtures. 
Case of a single beam configuration (left) with the two color beams
propagating in a coaxial fashion, and of a coplanar crossed-beam
dipole trap (right) with relative angle between the two beam pairs 
$\theta=0$. The lenses are assumed to be achromatic to ensure a 
common focus for the collimated beams at different wavelengths.}
\end{center}
\end{figure}

In the crossed-beam configuration, a pair of laser beams 
red-detuned with respect to the atomic transitions and focused 
on the center of the pre-existing trapping potential (for instance 
the one generated by the  magneto-optical trap typically 
used for precooling the atomic clouds), gives an effective 
attractive potential for both the species. 
This attractive potential is partially balanced by a second pair of 
mutually orthogonal blue-detuned laser beams, acting for instance 
along the same plane formed by the red-detuned beams and 
forming with the latter an angle $\theta$. 
This second pair of beams gives rise to a repulsive potential and, 
by a proper choice of its detuning and power, provides a selective 
deconfinement for the two species.

The potential generated by the dynamical Stark effect, felt
by an atom of species $\alpha$ ($\alpha=\mathrm{b}$ for bosons, 
$\alpha=\mathrm{f}$ for fermions) whose atomic transition wavelength 
and linewidth are respectively $\lambda_\alpha$ and $\Gamma_{\alpha}$, 
and due to the laser beams $i$ of wavelength $\lambda_i$ and
intensity $I_i$ ($i=1,2$ for the red-detuned and blue-detuned laser
beams, respectively), is \cite{ASKIN}
\begin{equation}
U^{\alpha}_i(x,y,z)=-\frac{\hbar \Gamma^2_{\alpha}}{8
I_{\alpha}^\mathrm{sat}} \left( {\frac{1}{\Omega_{\alpha}-\Omega_i}}+ 
{\frac{1}{\Omega_{\alpha}+\Omega_i}} \right) I_i(x,y,z),
\end{equation}
where $\Omega_{\alpha}=2\pi c/\lambda_{\alpha}$,
$\Omega_i=2\pi c/\lambda_i$, and $I_{\alpha}^\mathrm{sat}$ is
the saturation intensity for the atomic transition, 
expressed in terms of the former
quantities as $I_{\alpha}^\mathrm{sat}=\hbar \Omega_{\alpha}^3
\Gamma_{\alpha}/12 \pi c^2$. It is easy to recognize that the
potential energy has the same sign of the laser intensity if 
$\Omega_i > \Omega_\alpha$, \textit{i.e.} for blue-detuned light.
Then the atoms, trying to minimize their potential energy, move 
towards the regions of space with minimum light intensity, 
and therefore are expelled by the laser beam. The opposite 
occurs for red-detuned light, for which the atoms are attracted 
in the regions of maximum light intensity.

The laser intensity is the incoherent sum of the intensities
of each pair of beams (such incoherent sum can be obtained by 
orthogonal polarizations or by slight detuning of the beams 
within each pair), all focused at the origin of the trap. 
If we assume that the red-detuned beams propagate along the 
$x$-$y$ axes while the blue-detuned beams propagate along 
the $\xi$-$\eta$ axes possibly rotated by an angle $\theta$, 
\textit{i.e.} $\xi=x \cos\theta+ y \sin\theta$, 
$\eta=y \cos\theta-x\sin\theta$, with $0 \leq \theta<\pi/4$,
the intensities can be written as
\begin{eqnarray}
I_1(x,y,z) &=& {\frac{2 P_1}{\pi w_1^2}} \left\{
\frac{\exp\left[-\frac{2(y^2+z^2)}{w_1^2(1+ x^2/R_1^2)}\right]} 
{1+x^2/R_1^2} + 
\frac{\exp\left[-\frac{2( x^2+z^2)}{w_1^2(1+y^2/R_1^2)}\right]}
{1+y^2/R_1^2} \right\}
\label{intensity1}
\\
I_2(x,y,z)&=&{\frac{2 P_2}{\pi w_2^2}} \left\{
\frac{\exp\left[-\frac{2(\eta^2+z^2)}{w_2^2(1+ \xi^2/R_2^2)}\right]} 
{1+\xi^2/R_2^2} + 
\frac{\exp\left[-\frac{2( \xi^2+z^2)}{w_2^2(1+\eta^2/R_2^2)}\right]}
{1+\eta^2/R_2^2} \right\}
\label{intensity2}
\end{eqnarray}
where $P_i$ is the laser power, $w_i$ is the $1/e^2$ beam waist
radius, and $R_i=\pi w_i^2/\lambda_i$ the Rayleigh range. 
The quantity $2 P_i/\pi w_i^2=I_i(0,0,0)$ represents the peak laser
intensity due to each pair of beams, obtained in the focal point. 

The overall potential felt by the fermions (bosons) is 
$U_\mathrm{f}=U^\mathrm{f}_1+U^\mathrm{f}_2$ 
($U_\mathrm{b}=U^\mathrm{b}_1+U^\mathrm{b}_2$).
For a proper choice of the laser powers, these potentials are energy 
wells with depths $\Delta U_\mathrm{f}$ and $\Delta U_\mathrm{b}$ which 
constitute the confining energies of the fermionic and bosonic species.
The curvatures around the minimum of these potentials determine the 
trapping frequencies $\omega_\mathrm{f}$ and $\omega_\mathrm{b}$ and, 
consequently, the critical temperatures $T_\mathrm{F}$ and $T_\mathrm{c}$.
Up to quadratic terms, the potentials $U_\mathrm{f}$ and $U_\mathrm{b}$
are invariant under rotations around the minimum so that the
trapping frequencies do not depend on the angle $\theta$. 
By neglecting the terms $(\lambda_i/\pi w_i)^2 <<1$, we find
\begin{equation}
\label{omegas2}
\omega_{\alpha x}=\omega_{\alpha y}= 
\frac{\omega_{\alpha z}}{\sqrt{2}}= 
\sqrt{\frac{\hbar}{\pi m_{\alpha}} \left( 
\frac{k^{\alpha}_1 P_1}{w_1^4}+\frac{k^{\alpha}_2 P_2}{w_2^4} \right)},
\end{equation} 
where the constants $k^{\alpha}_i$ have been introduced as
\begin{equation}
k^{\alpha}_i=\frac{\Gamma^2_{\alpha}}{I_{\alpha}^\mathrm{sat}} \left( 
\frac{1}{\Omega_{\alpha}-\Omega_i}+
\frac{1}{\Omega_{\alpha}+\Omega_i} 
\right).
\end{equation}
Analogous expressions can be obtained in the case of a single-beam
geometry. For instance, by considering a single beam propagating
along the $x$-axis we have
\begin{equation}
\label{omegas1}
\omega_{\alpha x}= \sqrt{\frac{\hbar}{\pi m_{\alpha}} 
\left(\frac{k^{\alpha}_1P_1}{w_1^2 R_1^2}+
\frac{k^{\alpha}_2 P_2}{w_2^2 R_2^2} \right)},
\end{equation} 
\begin{equation}
\label{omegax1}
\omega_{\alpha z}=\omega_{\alpha y}= 
\sqrt{\frac{\hbar}{\pi m_{\alpha}} \left(\frac{k^{\alpha}_1
P_1}{w_1^4}+\frac{k^{\alpha}_2 P_2}{ w_2^4} \right)}.
\end{equation} 
Since typically $w_i \gg \lambda_i$ and, therefore, $R_i \gg w_i$, 
the confinement along the $x$-direction is weakened leading, 
with respect to the crossed-beam configuration, to an overall 
reduction of the average angular frequencies for the trapped species. 
In the particular case of equal ratios between wavelengths and 
waists for the two colors, $\lambda_1/w_1=\lambda_2/w_2=\lambda/w$, 
this reduction amounts to $(\lambda/2\pi w)^{1/3}$.

In Figure 2 we show the dependence of the optical potential 
upon the radial, $x$ or $y$, and axial, $z$, directions in 
the two-beam geometry for the case of the $^6$Li-$^{23}$Na mixture.
\begin{figure}
\begin{center}
\psfrag{x}[c]{$x$} \psfrag{yx}[c][l][1]{$U_\mathrm{f/b}(x,0,0)$ (mK)}
\psfrag{z}[c]{$z$} \psfrag{yz}[c][l][1]{$U_\mathrm{f/b}(0,0,z)$ (mK)}
\psfrag{Na}{$^{23}$Na} \psfrag{Li}{$^{6}$Li}
\includegraphics[width=1.0\textwidth]{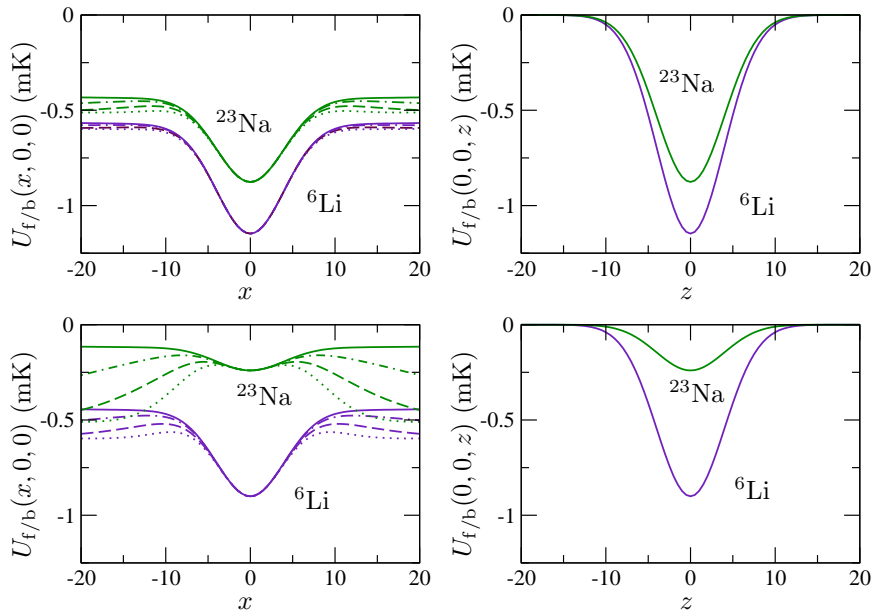}
\caption{Potential energies in the case of the $^6$Li-$^{23}$Na
mixture in a bichromatic optical dipole trap in a crossed-beam
geometry formed by focusing a Nd:YAG laser beam ($\lambda_1=$1064 nm) 
and a blue-detuned laser beam at its second harmonic
($\lambda_2=$532 nm). 
In the two upper panels the potential energy for the atoms
of the two species is shown along the radial, $x$ or $y$, 
and axial, $z$, directions for a power ratio $P_2/P_1=0.05$. 
The four curves in the radial directions refer 
to various angles ($\theta=0$, $\pi/16$, $\pi/8$, and $\pi/4$
from top to bottom) between the two pairs of beams.
In the two lower panels the case of a larger power ratio, 
$P_2/P_1=0.25$, is considered. 
The beam waists are $w_1=w_2=8$ $\mu$m, and $P_1=1$ W.
Note that while the potential vanishes with a purely Gaussian
behaviour in the axial direction $z$, along $x$ (and $y$) first it 
varies in a Gaussian way (to half of its peak value for $\theta=0$)
and finally vanishes as a Lorentzian (not shown). 
The Gaussian and Lorentzian widths are determined by the beam waists 
and the Rayleigh ranges, respectively.}
\end{center}
\end{figure}
When the ratio between the power of the blue- and red-detuned 
lasers is small ($P_2/P_1=0.05$, upper panels) the deformations
induced by the blue-detuned beam are negligible. 
The difference in curvature between the two species is mainly
attributable to the difference in mass $m_{\alpha}$ and detuning with
respect to the laser beam, both playing a role in favoring 
$T_\mathrm{c}$ smaller than $T_\mathrm{F}$.  
Also, the depth of the potential well for the bosonic species 
is smaller than that for the fermions. 
This makes possible to exploit evaporative cooling 
without appreciable interference from the Fermi cloud. 
It is also evident that the stronger confinement in the radial 
direction is achieved for coaxial beams, \textit{i.e.} $\theta=0$. 
The axial confinement along the $z$ axis is instead unaffected 
by the rotation angle between the beam pairs.
In the case of a strong perturbation, lower panels of Fig. 2 
where $P_2/P_1=0.25$, the bosonic species is much less confined 
and there is a strong difference in the curvature of the energy
potential with respect to the fermionic species. 

In both crossed- or single-beam configurations, the confinement 
energy of the boson species vanishes as the power ratio $P_2/P_1$ 
approaches a critical value. Approximately, this can be obtained 
by Eq. (\ref{omegas2}) or Eqs. (\ref{omegas1}-\ref{omegax1}) as 
the $P_2/P_1$ value at which the average trapping frequency of bosons vanishes.
In both cases, we find
\begin{equation}
\label{p2p1crit}
\left. \frac{P_2}{P_1} \right|_{\mathrm{crit}}= 
{\frac{\Omega_2^2-\Omega_\mathrm{b}^2}{\Omega_\mathrm{b}^2-\Omega_1^2}} 
{\left(\frac{w_2}{w_1}\right)}^4.
\end{equation}

The behavior of the trapping frequency ratio 
$\omega_\mathrm{f}/\omega_\mathrm{b}$ as well as of the trapping
energies $\Delta U_\mathrm{f}$ and $\Delta U_\mathrm{b}$ 
as a function of the power ratio $P_2/P_1$ is shown
in Figs. 3 and 4 also for different Fermi-Bose mixtures, 
obtained by using fermionic $^{40}$K and bosonic
$^{87}$Rb, as summarized in Table 2. 
\begin{table}
\begin{center}
\begin{tabular}{cccccc}
\hline\hline
mixture & $\lambda_\mathrm{f}$ (nm) & $\Gamma_\mathrm{f}$ (MHz) & $\lambda_\mathrm{b}$ (nm) & $\Gamma_\mathrm{b}$ (MHz) & $\lambda_2$ (nm) \\
\hline
$^{6}$Li-$^{23}$Na & 671 & $2\pi\times 5.9 \phantom{0}$ & 589 & $2\pi\times 9.8 \phantom{0}$ & 532$\phantom{.0}$ \\
$^{6}$Li-$^{87}$Rb & 671 & $2\pi\times 5.9 \phantom{0}$ & 780 & $2\pi\times 5.98$ & 740$\phantom{.0}$ \\
$^{40}$K-$^{23}$Na & 767 & $2\pi\times 6.09$ & 589 & $2\pi\times 9.8 \phantom{0}$ & 532$\phantom{.0}$ \\
$^{40}$K-$^{87}$Rb & 767 & $2\pi\times 6.09$ & 780 & $2\pi\times 5.98$ & 773.5 \\
\hline\hline
\end{tabular}
\end{center}
\caption{Wavelenghts and linewidths of the atomic transitions for
the fermion-boson mixtures considered in the text and wavelength 
of the corresponding deconfining laser with a Nd:YAG laser at 
$\lambda_1=1064$ nm producing the primary trapping potential.}
\end{table}
\begin{figure}
\begin{center}
\psfrag{Na-Li}{$^{6}$Li-$^{23}$Na} \psfrag{Rb-Li}{$^{6}$Li-$^{87}$Rb}
\psfrag{Na-K}{$^{40}$K-$^{23}$Na} \psfrag{Rb-K}{$^{40}$K-$^{87}$Rb}
\psfrag{x}[]{$P_2/P_1$}
\psfrag{y}[]{$\omega_\mathrm{f}/\omega_\mathrm{b}$}
\includegraphics[width=1.0\textwidth]{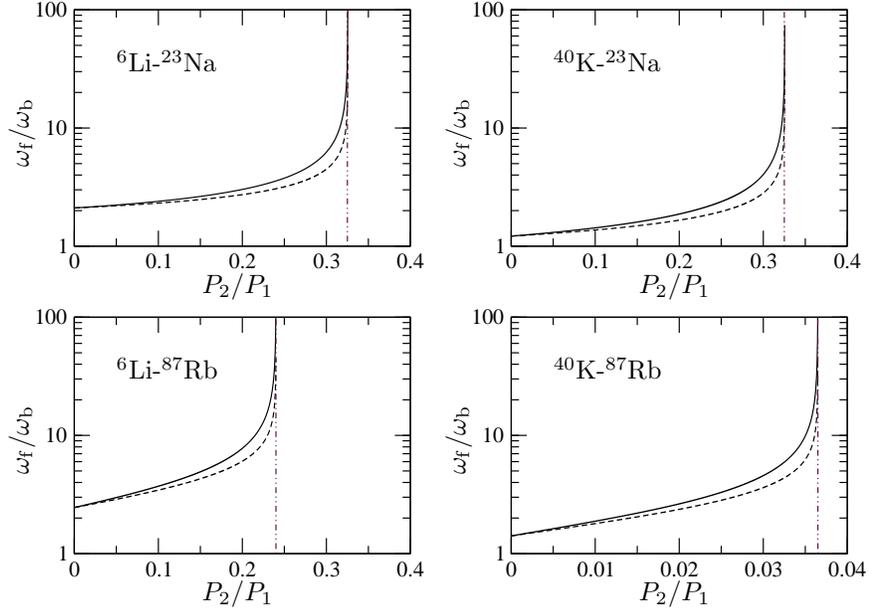}
\caption{ Selective trapping of bosons and fermions. The ratio
between the average trapping frequencies for the fermionic and
bosonic species is shown versus the beam power ratio between the
blue- and red-detuned lasers. The cases of single beam (dashed lines)
and crossed beam (solid lines) geometries are depicted for the 
Li-Na, Li-Rb, K-Na, and K-Rb mixtures. The vertical lines are the 
critical values $P_2/P_1$  given by Eq. (\ref{p2p1crit}).
Parameters chosen according to the values in Table 2.}
\end{center}
\end{figure}
\begin{figure}
\begin{center} 
\psfrag{Na}{$^{23}$Na} 
\psfrag{Li}{$^{6}$Li} 
\psfrag{K}{$^{40}$K}
\psfrag{Rb}{$^{87}$Rb} 
\psfrag{x}[]{$P_2/P_1$} 
\psfrag{y}[]{$\Delta U/P_1$ (mK/W)} 
\includegraphics[width=1.0\textwidth]{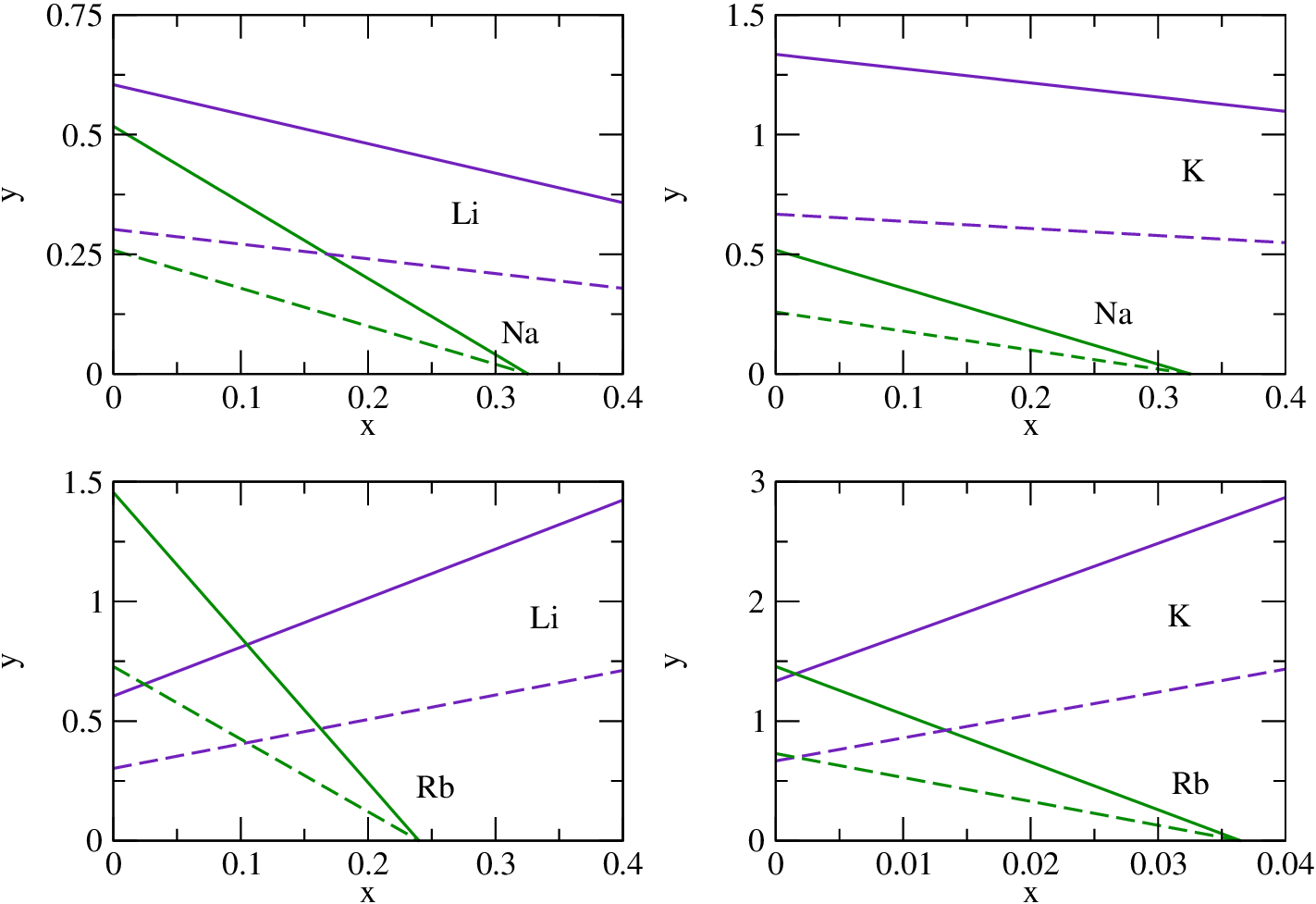}
\caption{Selective trapping of bosons and fermions. 
Confining energy per unit of infrared laser power as a function
of the beam power ratio between the blue- and red-detuned lasers.
The confining energy is evaluated with respect to the region of space 
delimited by the waist size (see also note in the caption of Figure 2). 
Continuous lines refer to the crossed-beam configuration, while
dashed lines are for the single-beam optical dipole trap.}
\end{center}
\end{figure}
The trapping frequency ratio for the single-beam geometry is always 
reduced with respect to the corresponding crossed-beam case at the
same power ratio $P_2/P_1$.
This decrease, which is a consequence of the weaker confinement along 
the direction of the laser beam, can be compensated by a larger power 
ratio, provided that the beam intensities are adequately stabilized. 
From the upper panels in Figure 4 we see that, with the chosen values 
of the blue-detuned laser wavelength, the bosonic sodium atoms are 
always less confined than the fermionic atoms. This allows one 
to apply evaporative cooling techniques for the bosonic species 
while having negligible losses in the number of fermions. 
For the mixtures utilizing $^{87}$Rb, the confining energy is
initially higher for the Bose species (especially for the Li-Rb
mixtures), however the fermionic confining energy increases as the 
laser power $P_2$ is increased, since the potential created by 
the blue-detuned light is in this case attractive for fermions 
($\lambda_2 > \lambda_\mathrm{f}$).  
For these mixtures, therefore, the addition of the laser light 
deconfining the bosons improves at the same time the confinement 
features of the fermionic species and minimizes the losses of 
Fermi atoms at the beginning of evaporation.
Note also that for all the mixtures considered, within a very good 
approximation the bosonic confinement energies $\Delta U_\mathrm{b}$ in
Figure 4 vanish at the same critical values $P_2/P_1$ at which 
the ratios $\omega_\mathrm{f}/\omega_\mathrm{b}$ in Figure 3 diverge.

Some further comments are in order. 
In the case of equal waists, the potential felt by the bosons 
is exactly zero at the critical power ratio and changes curvature 
at larger values giving rise to deconfinement, see middle panel
of Figure 5. 
\begin{figure}
\begin{center}
\psfrag{z}{$z$} 
\psfrag{U_b(0,0,z) (mK)}{$U_\mathrm{b}(0,0,z)$ (mK)}
\psfrag{w2/w1=0.9}{$w_2/w_1=0.9$}
\psfrag{w2/w1=1}{$w_2/w_1=1$}
\psfrag{w2/w1=1.1}{$w_2/w_1=1.1$}
\includegraphics[width=0.7\textwidth]{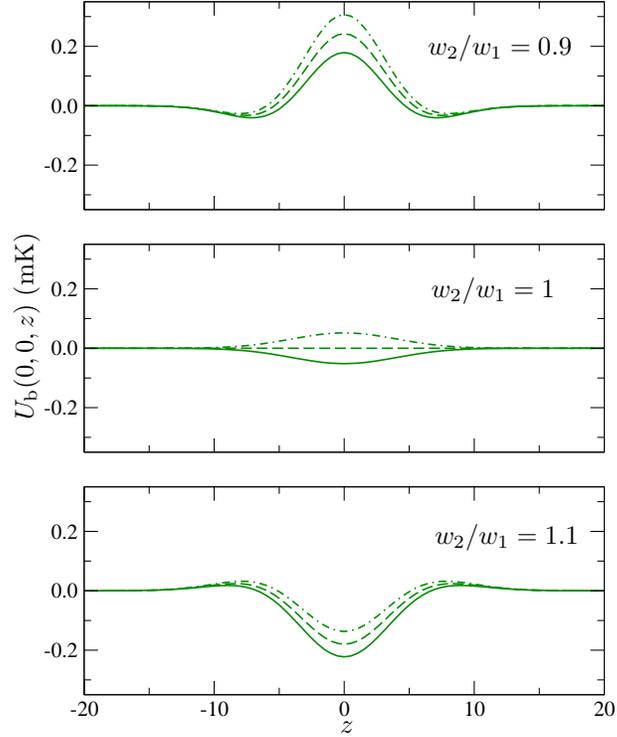}
\caption{Bosonic potential energy along the $z$-axis 
in the case of the $^6$Li-$^{23}$Na mixture in a crossed-beam 
bichromatic optical dipole trap
for $P_2/P_1$ 5\% smaller (solid), equal (dashed), and
5\% larger (dot-dashed) than ${P_2 / P_1}|_{\mathrm{crit}}$.
The three panels refer to waist ratios $w_2/w_1=0.9$ (top),
$w_2/w_1=1$ (middle), and $w_2/w_1=1.1$ (bottom).}
\end{center}
\end{figure}
By choosing $w_2 < w_1$ (top panel of Figure 5), 
the potential becomes bistable for $P_2/P_1$ around the critical
value. Then, the blue-detuned laser can be used to produce tailored 
bistable potentials, providing an alternative route to the recently 
demonstrated all-magnetic bistable potentials \cite{THOMAS}.
For $w_2 > w_1$ (bottom panel of Figure 5), the potential 
is always globally deconfining but a local minimum around the origin is
assured if $P_2/P_1$ is close to the critical value. 
The strong dependence of the critical power ratio upon the 
beam waists can be used to reduce the amount of blue-detuned light
necessary to approach the targeted values of 
$\omega_{\mathrm{f}}/\omega_{\mathrm{b}}$. 
However, based on the above mentioned considerations, there is a 
tradeoff since we may have also a change of the potential shape.

A possible technical issue is the demand for an initial large laser
power especially since it is difficult to get sub-Doppler cooling for
Li and K due to the small hyperfine splittings of the $p_{3/2}$
atomic level \cite{SCHUENEMANN} (see however \cite{MODUGNO1} for 
a successful demonstration of sub-Doppler cooling of $^{40}$K). 
Up to now, Nd:YAG laser powers with built-in crystals for 
frequency-doubling are limited to about 1 W.
This corresponds to order of 500 $\mu$K for the initial potential depth, 
\textit{i.e.} four times the initial temperature of the cloud if
the latter is transferred from a Doppler-limited magneto-optical trap.
As we will discuss in the next Section, this could be an issue 
for starting efficient evaporative cooling. 
Besides awaiting progress in the power deliverable by 
semiconductor-based lasers, one can use 
independent systems for red- and blue-detuned light (this seems 
anyway unavoidable for mixtures utilizing $^{87}$Rb as the Bose cooler).
High power, far detuned CO$_2$ lasers are also available to produce 
quasi-static optical dipole traps \cite{TAKEKOSHI1,TAKEKOSHI2}.
This technique has been recently demonstrated for cooling 
fermionic lithium at degeneracy temperature \cite{GRANADE} after 
efficient loading from a magneto-optical trap \cite{OHARA,OHARA1}. 
CO$_2$ lasers allow for large powers (order of 100-200 W) and, 
since they are well detuned from the atomic resonances,  
for very small heating due to photon scattering \cite{SAVARD}. 
In Figure 6 we show the selective trapping features of a K-Na mixture
with the Nd:YAG laser replaced by a CO$_2$ laser emitting at 10.6 $\mu$m. 
\begin{figure}
\begin{center}
\psfrag{Na}{$^{23}$Na} \psfrag{K}{$^{40}$K} 
\psfrag{x}[]{$P_2/P_1$}
\psfrag{y1}[]{$\omega_\mathrm{f}/\omega_\mathrm{b}$}
\psfrag{y2}[]{$\Delta U/P_1$ ($\mu$K/W)}
\includegraphics[width=1.0\textwidth]{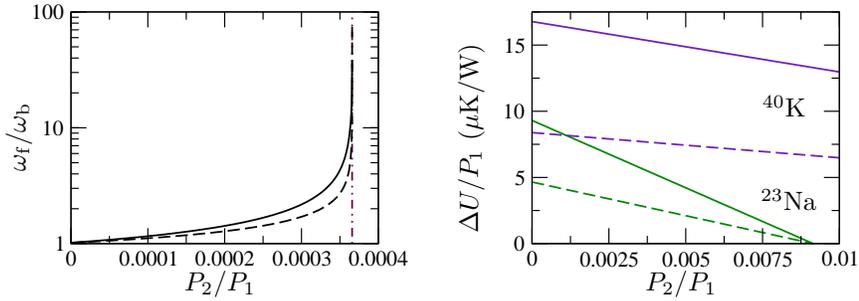}
\caption{Selective trapping with a CO$_2$ laser in the case of the
K-Na mixture. Trapping frequency ratio (left) and confinement energy
(right) versus the parameter $P_2/P_1$. The dashed curves refer to
the single-beam configuration, the continuous ones to the crossed-beam 
geometry. The different critical values of $P_2/P_1$ obtained from the 
left and right panels are due to the fact that the harmonic
approximation used to evaluate $\omega_\mathrm{f}/\omega_\mathrm{b}$ 
fails for $P_2/P_1$ large enough (the potential becomes bistable 
since $w_2 < w_1$). The waist of the CO$_2$ laser beam is assumed 
to be $w_1=50$ $\mu$m, while $w_2=10$ $\mu$m.}
\end{center}
\end{figure}
At such large wavelengths the optical potential is well approximated 
by the induced dipole interaction $U=-\alpha_\mathrm{g} \overline{E^2}/2$, 
where $\alpha_\mathrm{g}$ is the ground state polarizability and 
$\overline{E^2}$ is the time-averaged squared electric field 
\cite{TAKEKOSHI1}. 
In terms of the laser intensity the optical potential is
\begin{equation}
U(x,y,z)=-\frac{2\pi \alpha_\mathrm{g}}{c} ~I(x,y,z)
\end{equation}
Due to the high static atomic polarizability of rubidium
($\alpha_\mathrm{g}^\mathrm{Rb}=47.3 \times 10^{-24}$ cm$^3$), 
this bosonic species is more strongly confined than fermionic
potassium ($\alpha_\mathrm{g}^\mathrm{K}=43.4 \times 10^{-24}$ cm$^3$)
and lithium ($\alpha_\mathrm{g}^\mathrm{Li}=24.3 \times 10^{-24}$ cm$^3$), 
therefore ruling out evaporation as a possible cooling technique
for K-Rb and Li-Rb mixtures. Also, the combination Li-Na has to be 
discarded due to the small static atomic polarizability of sodium
($\alpha_\mathrm{g}^\mathrm{Na}=24.08 \times 10^{-24}$ cm$^3$). 
Thus, a CO$_2$ laser can be effectively used for a red-detuned 
attractive potential only in the case of the $^{40}$K-$^{23}$Na 
mixture considered in Figure 6.

\section{Evaporative cooling}

So far we have considered the static trapping features of the
two-color optical dipole configuration schematized through a 
time-independent potential. However, to reach quantum degeneracy, 
the phase space density has to be increased, \textit{e.g.} 
by cooling down the atomic sample. In the configuration proposed here 
this is obtained through two continuous processes: forced and
selective removal of bosonic atoms with re-thermalization
of the surviving component (evaporative cooling) \cite{KETTERLE}, 
and thermalization of the fermionic species to the temperature of the 
Bose gas (sympathetic cooling). These cooling processes have various 
limitations, the most obvious one being the reduced heat capacity of 
the bosonic sample undergoing a continuous decrease of atoms. 
Moreover, concurrent heating sources will limit the ultimate reachable
temperature, most notably the residual Rayleigh scattering from the
trapping beams. In this Section we discuss to some extent these issues and, 
although not pretending to go into full details on the many possible 
experimental schemes, we give estimates of the relevant parameters 
involved in the cooling dynamics for a specific example. 

Evaporative cooling in an optical dipole trap has been the subject of 
various experimental and theoretical studies \cite{ADAMS,BARRETT,OHARA2}. 
The atomic evaporation rate in finite-depth traps is exponentially
dependent upon the ratio between the potential depth of the trap and 
the average thermal energy of the atomic cloud, $\Delta U/k_\mathrm{B} T$,
since the number of atoms in the tail of the Boltzmann distribution 
scales as $\exp(-\Delta U/k_\mathrm{B} T)$. 
Forced evaporation is, therefore, necessary to maintain a significant 
evaporation rate, which should be otherwise exponentially quenched 
by reducing the cloud temperature. 
It can be demonstrated \cite{OHARA2} that all the relevant quantities 
involved in forced evaporative cooling in an optical dipole trap 
scale with some power of the confining energy, 
provided that the ratio between the latter and the cloud temperature
is constant in time $\Delta U/k_\mathrm{B} T=\eta$. 
In this case, the confining energy has a time dependence of the form
\begin{equation}
\frac{\Delta U(t)}{\Delta U_\mathrm{i}}=
{\left(1 + \frac{t}{\tau} \right)}^{\varepsilon_U},
\end{equation}
where $\Delta U_\mathrm{i}$ is the initial potential depth, 
$\varepsilon_U=-2(\eta^\prime-3)/\eta^\prime$, and 
$\tau^{-1}= (2/3) \eta^\prime (\eta-4) \exp(-\eta)\gamma_\mathrm{i}$, 
with $\eta^\prime=\eta+(\eta-5)/(\eta-4)$ and $\gamma_\mathrm{i}$ 
being the initial value of the elastic collision rate $\gamma$. 
Similar laws hold for the time dependence of the number of particles $N$, 
temperature $T$, phase-space density $\rho$, and elastic collision
rate $\gamma$ with corresponding exponents 
$\varepsilon_N$, $\varepsilon_T$, $\varepsilon_\rho$, and
$\varepsilon_{\gamma}$.

With respect to the simplest case of a single Bose species undergoing 
evaporative cooling in an optical dipole trap we have two differences. 
First, there is a mixture also containing the Fermi atoms. 
Their presence does not significantly affect evaporative 
cooling of the Bose gas if the potential energy depth 
$\Delta U_\mathrm{f}$ is much larger than $\Delta U_\mathrm{b}$, a 
condition well satisfied in the situations we consider - see Figure 4. 
Also, in order for the Bose gas to act as a cooler its heat capacity 
must be much larger than the heat capacity of the fermionic species.
This is possible, at least in the earlier stages of evaporation 
before entering the degenerate regime, by assuming a number of 
bosons much larger than the number of fermions.   
In bichromatic traps, a further problem may arise due to the presence 
of the blue-detuned laser which weaken the confinement of both species, 
particularly the bosonic one. As the atomic densities are decreased, 
all the density-dependent scattering rates, in particular the elastic 
one which is crucial for thermalization of the surviving atoms, 
are also suppressed \cite{HOUBIERS}. 

The lower densities and the slower thermalization at lower 
temperatures makes more difficult to achieve high phase-space densities, 
an issue which has limited for various years the effort to make an 
all-optical formation of Bose-Einstein condensates \cite{NOTE1}.  
From this point of view, in a bichromatic trap it is convenient to 
turn on the blue-detuned beam only when it can give an effective 
advantage, \textit{i.e.} close to the Fermi degenerate regime. 
Once on, the power of the blue-detuned beam must be chosen in such
a way to maintain a fixed ratio with respect to that of the red-detuned 
beam. Meanwhile the power of the red-detuned beam is decreased 
continuously to allow efficient forced evaporation of the bosonic 
species. An example of this cooling strategy is shown in Figure 7.
\begin{figure}
\begin{center}
\psfrag{P_1}{$P_1$}
\psfrag{P_2}{$P_2$}
\psfrag{w_f}{$\omega_\mathrm{f}$}
\psfrag{w_b}{$\omega_\mathrm{b}$}
\psfrag{time}{$\gamma_\mathrm{i} t$}
\psfrag{powers}{$P$ (W)}
\psfrag{trap frequencies}{$\omega/2\pi$ (kHz)}
\psfrag{elastic rate}{$\gamma$ (Hz)}
\includegraphics[width=1.0\textwidth]{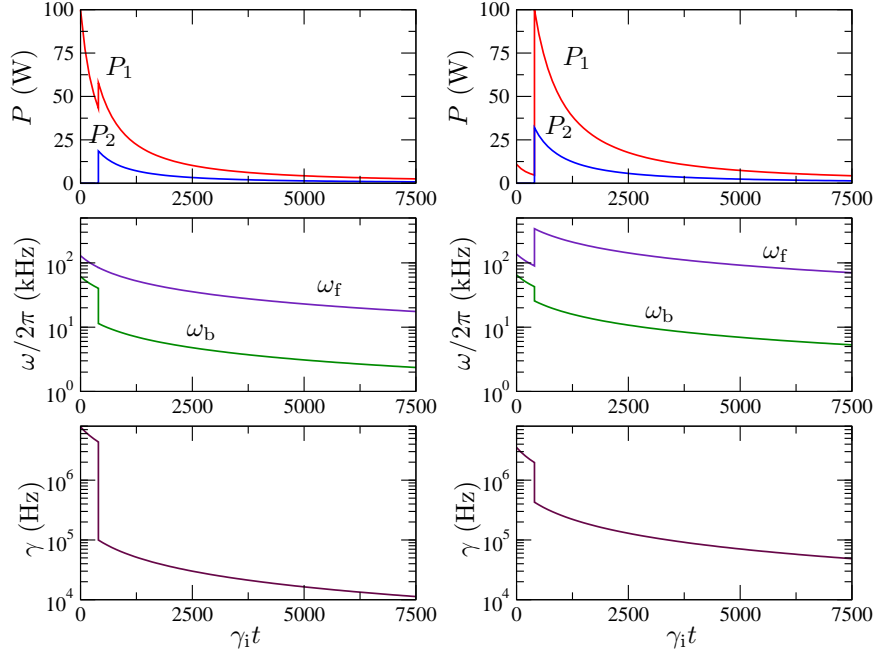}
\caption{Evaporative cooling strategies in bichromatic optical 
dipole traps. The time dependence of the laser powers (top panels), 
the averaged frequencies (center panels), and the bosonic elastic
scattering rates (bottom panels), are depicted for single-beam (left) 
and crossed-beam (right) configurations and a $^6$Li-$^{23}$Na mixture. 
The red-detuned beam is obtained with a Nd:YAG laser emitting at 
$\lambda_1=1064$ nm with a peak power $P_1$=100 W. 
For $\gamma_\mathrm{i} t >400$, the ratio between the blue- and red-detuned 
laser powers is kept at the constant value $P_2/P_1=0.32$. 
In the case of the single-beam configuration there is a significant 
decrease in the bosonic trapping frequency, while in the crossed-beam 
configuration a further gain through the increase of the fermionic 
trapping frequency is also evident. 
The large elastic scattering rates available during the entire 
evaporation process allows for a fast dynamics of thermalization 
of the Bose gas. We assume equal waists for the beams, $w_1=w_2=8$ $\mu$m. 
The initial laser power is enough to have $\eta=10$ for a $^{23}$Na
cloud consisting of $N_\mathrm{b}=10^6$ atoms transferred from a 
magneto-optical trap at temperature $T=586$ $\mu$K (crossed-beam) or 
$T=222$ $\mu$K (single-beam), thermalized with a $^6$Li cloud with 
$N_\mathrm{f}=10^5$ atoms.}
\label{fig:fig7}
\end{center}
\end{figure}

The constant power ratio $P_2/P_1$ must be chosen carefully 
as a compromise between increasing the fermion-to-boson trapping 
frequency ratio and not decreasing too much the absolute frequencies, 
as this will decrease all the elastic scattering rates crucial for 
interspecies and intraspecies thermalization. 
In fact, the elastic collision rate for indistinguishable bosons is given by
\begin{equation}
\gamma=
 N_\mathrm{b} m_\mathrm{b} 
\omega_\mathrm{b}^3 \sigma_\mathrm{b}/2 \pi^2 k_\mathrm{B} T,
\end{equation} 
where $\sigma_\mathrm{b} = 8 \pi a_\mathrm{b}^2$ is the boson elastic 
cross-section expressed in terms of the boson $s$-wave elastic
scattering length $a_\mathrm{b}$. The value of $\gamma$ during 
the cooling strategies described here is shown in the bottom panels 
of Figure 7.

If the technical noise and heating sources are properly reduced 
and the residual background pressure is low enough, the ultimate 
heating source is set by residual Rayleigh scattering from the laser 
beams. 
This depends on various parameters, most notably the laser intensity 
and the detuning of the laser beam from the atomic resonance, and 
the corresponding rate reads
\begin{equation}
\gamma_i^\alpha(x,y,z)= \frac{\Gamma_\alpha^3}{8} 
\left( \frac{\Omega_i}{\Omega_{\alpha}} \right)^3 
\left( \frac{1} {\Omega_{\alpha}-\Omega_i}+ 
\frac{1}{\Omega_{\alpha}+\Omega_i} \right)^2 
\frac{I(x,y,z)}{I_{\alpha}^\mathrm{sat}}.
\end{equation}
In the following we will estimate the peak Rayleigh scattering rate,
\textit{i.e.} the value corresponding to the peak laser intensity
in the center of the trapping potential, per unit of power of the 
laser beams in the case of the $^{6}$Li-$^{23}$Na mixture used 
in \cite{HADZIBABIC}. 
The time dependence of the Rayleigh scattering rate can then be inferred 
by properly scaling the curves for the laser powers shown in Figure 7. 
As expected, the strongest source of heating comes from 
the blue-detuned laser beam acting on the sodium atoms. 
This is of less concern because the blue-detuned laser beam 
is turned on only during the latest stage of evaporation, and 
then raised to a fraction of the red-detuned laser beam. 
The residual Rayleigh scattering rate is estimated to be around 
$1.2 \times 10^{-2}$ Hz, much smaller than the estimated 
elastic collisional rate, and corresponds to lifetimes well in excess of 10 s. 
A similar situation occurs also for the $^{40}$K-$^{23}$Na mixture. 
As shown in Table 3, due to the proximity of the atomic transitions 
for $^6$Li or $^{40}$K and $^{87}$Rb instead, the Rayleigh scattering 
rates due to the deconfining beam seems prohibitive.
\begin{table}
\begin{center}
\begin{tabular}{ccccc}
\hline\hline
mixture & $\gamma_1^\mathrm{f}$ (Hz) & $\gamma_2^\mathrm{f}$ (Hz) 
& $\gamma_1^\mathrm{b}$ (Hz) & $\gamma_2^\mathrm{b}$ (Hz) \\
\hline
$^{6}$Li-$^{23}$Na & $0.94$ & $12.5$ & 0.8 & $63$ \\
$^{6}$Li-$^{87}$Rb & $0.94$ & $31$ & 4.7 & $306$ \\
$^{40}$K-$^{23}$Na & $4.7$ & $7.8$ & 0.8 & $63$ \\
$^{40}$K-$^{87}$Rb & $4.7$ & $9.4 \times 10^{3}$ & 4.7 & $1.1 \times 10^4$\\
\hline\hline
\end{tabular}
\end{center}
\caption{Estimate of the Rayleigh scattering rates per unit of the 
corresponding laser power for the fermion-boson mixtures considered 
in the text. The laser wavelengths are the same as in Table 2  
and the beam waists are chosen as $w_1=w_2=8$ $\mu$m.}
\end{table} 
However, the large Rayleigh scattering rate per unit of power for the 
$^{40}$K-$^{87}$Rb mixture is less frightening since this mixture 
has also a critical power ratio smaller by one order of magnitude 
with respect to the other ones, if equal waists are assumed. 

The lifetime estimated for $^{6}$Li-$^{23}$Na is long enough 
to perform various experiments aimed at evidencing superfluidity 
features, such as mechanical stirring or generation of collective 
excitations. 

\section{Sympathetic cooling and possible evidences for a fermionic 
superfluid phase}

The study of the dynamics of evaporative cooling is a prerequisite to
discuss sympathetic cooling of fermions through their interactions
with the Bose gas. 
Heat exchange between two ensembles is perhaps the most widespread 
thermodynamic process occurring in nature. 
Thus, as a general method to refrigerate an atomic or molecular 
ensemble whose direct cooling is difficult to achieve, 
one can spatially and temporally overlap it with a colder ensemble, 
the so-called sympathetic cooling. 
This process has been first demonstrated at the microscopic level 
for trapped ions \cite{LARSON}, then for neutral atoms and molecules 
via use of cryogenically cooled helium gas \cite{DECARVALHO}. 
More recently, with the advent of ultracold atomic physics, 
sympathetic cooling in the microkelvin and nanokelvin ranges 
has been achieved for bosons in different internal states \cite{MYATT}, 
in different isotopes \cite{BLOCH}, and for different species \cite{MODUGNO}. 

The elastic scattering properties among fermions and bosons at 
very low temperatures, which strongly influence the efficacy 
of sympathetic cooling, are starting to be collected for
various atomic mixtures \cite{HADZIBABIC,ROATI}. 
One potential problem common to all the mixtures is the diminished 
heat exchange capability of bosons when approaching condensation.
In particular, fermions can be considered as impurities in the boson 
cloud and below $T_\mathrm{c}$ the Bose gas has a condensed fraction
which is expected to be superfluid. 
Based on the Landau criterion for the critical velocity in a superfluid, 
we do expect Fermi-Bose collisions to be suppressed when the Fermi 
velocity becomes smaller than the sound velocity of the Bose gas
\cite{TIMMERMANS}. 
This has been also experimentally demonstrated \cite{CHIKKATUR}. 
The Fermi velocity, defined by  $k_\mathrm{B} T_\mathrm{F}=
m_\mathrm{f} v_\mathrm{F}^2/2$, can be expressed in terms of the 
trapping parameters as
\begin{equation}
v_\mathrm{F}=1.91 
\left(\frac{\hbar \omega_\mathrm{f}}{m_\mathrm{f}}\right)^{1/2} 
N_\mathrm{f}^{1/6}.
\end{equation}  
On the other hand, the sound velocity $v_\mathrm{s}$ for the Bose 
gas can be written in terms of the chemical potential as 
$\mu=m_\mathrm{b} v_\mathrm{s}^2$leading, in the Thomas-Fermi limit, to
\begin{equation}
v_\mathrm{s}=1.22
\left(\frac{\hbar^2 \omega_\mathrm{b}^3 a_\mathrm{b} N_\mathrm{b}}
{m_\mathrm{b}^2}\right)^{1/5}, 
\end{equation}
where $a_\mathrm{b}$ is the boson $s$-wave scattering length. 
The $v_\mathrm{F}/v_\mathrm{s}$ ratio is, therefore,
\begin{equation}
\frac{v_\mathrm{F}}{v_\mathrm{s}} = 1.57 
\left(\frac{\hbar}{m_\mathrm{b} \omega_\mathrm{b} 
a_\mathrm{b}^2} \right)^{1/10}
\left(\frac{m_\mathrm{b}}{m_\mathrm{f}}\right)^{1/2}
\left(\frac{\omega_\mathrm{f}}{\omega_\mathrm{b}}\right)^{1/2}
\frac{N_\mathrm{f}^{1/6}}{N_\mathrm{b}^{1/5}}.
\label{vratio}
\end{equation}

The following remarks are in order: 
\textit{a}) the velocity ratio (\ref{vratio}) scales with the 
square root of the trapping frequencies, so that it helps to have
$\omega_\mathrm{f}/\omega_\mathrm{b} >>1$ 
to avoid suppression of Fermi-Bose scattering;
\textit{b}) the velocity ratio also depends upon the number of 
fermions and bosons. If the former are kept constant in the trap and 
the latter undergo evaporative cooling, the velocity ratio is increased;
\textit{c}) the ratio $v_\mathrm{F}/v_\mathrm{s}$ is already 
large for conventional single-color optical dipole traps. 
For instance, if $\omega_\mathrm{b}= 2\pi \times 10^4$ s$^{-1}$ and 
$N_\mathrm{f}=N_\mathrm{b}$, for a $^{6}$Li-$^{23}$Na mixture we have 
$v_\mathrm{F}/v_\mathrm{s} \simeq 7.2$. 
In this case, the loss of cooling efficiency becomes relevant 
for fermion velocities $v \lesssim v_\mathrm{F} /7.2$ 
corresponding to $T/T_\mathrm{F} \lesssim 2 \times 10^{-2}$. 
It seems therefore that this mixture will hardly enter into the regime 
where superfluid suppression of impurity scattering is significant. 

A more stringent limitation to sympathetic cooling is instead set 
by the classical heat exchange between bosons and fermions. 
In Figure 8 we show the dependence of the ratio between the heat capacities
for the Bose and the Fermi species, $C_\mathrm{b}/C_\mathrm{f}$, 
in bichromatic traps with single and crossed-beam configurations. 
\begin{figure}
\begin{center}
\psfrag{T/T_F}{$T/T_\mathrm{F}$}
\psfrag{T/T_c}{$T/T_\mathrm{c}$}
\psfrag{C_b/C_f}{$C_\mathrm{b}/C_\mathrm{f}$}
\psfrag{T/T_F T/T_c}{$T/T_\mathrm{F}$, $T/T_\mathrm{c}$}
\psfrag{time}{$\gamma_\mathrm{i} t$}
\includegraphics[width=0.7\textwidth]{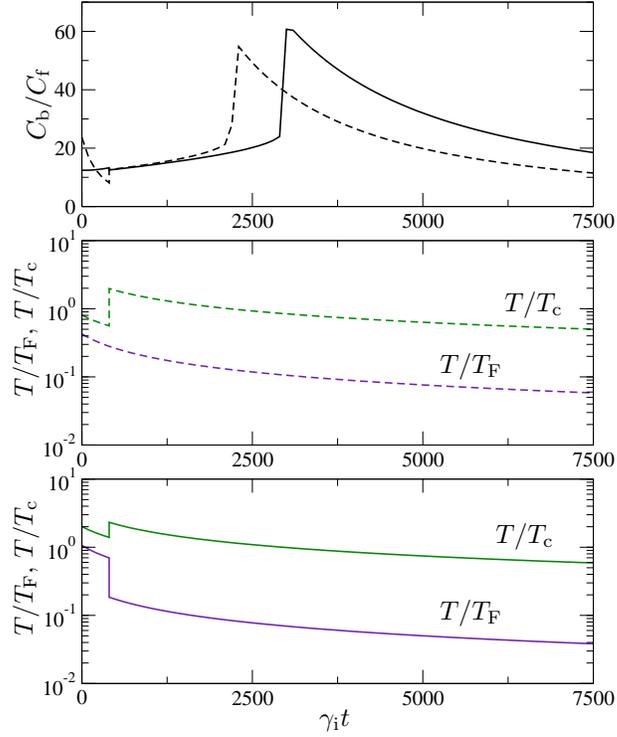}
\caption{
Efficiency for sympathetic cooling of Fermi-Bose mixtures in 
bichromatic optical dipole traps. In the top panel the ratio between 
the heat capacities of bosons and fermions $C_\mathrm{b}/C_\mathrm{f}$ 
is shown versus time for the evaporative cooling strategies chosen
in Figure 7 for single-beam (dashed) and crossed-beam (solid) 
configurations. The time evolution of the temperature ratios 
$T/T_\mathrm{F}$ and $T/T_\mathrm{c}$ is shown for the single-beam 
(center panel) and crossed-beam (bottom  panel) configurations.} 
\label{fig:fig8}
\end{center}
\end{figure}
This gives us a picture of the efficiency of sympathetic cooling since 
this process breaks down when the ratio $C_\mathrm{b}/C_\mathrm{f}$
becomes of the order of unity. 
The heat capacities have been evaluated numerically as described in
Appendix A taking into account the time dependence of the various 
parameters, in particular the diminishing number of bosons during
forced evaporation. 
The effect of many-body interactions on the heat capacity of bosons, 
evaluated in \cite{GIORGINI}, is next-to-leading with respect to the 
dependence upon the number of bosons. 
We see that in the bichromatic traps considered here
the heat capacity of the bosons is always larger than that of 
the fermions by an order of magnitude.
 
In Figure 8 we also show the temperature ratios $T/T_\mathrm{F}$
and $T/T_\mathrm{c}$. In both single- and double-beam cases, it is 
possible to reach $T/T_\mathrm{F} < 10^{-1}$ while for the Bose gas  
$T/T_\mathrm{c} > 5 \times 10^{-1}$, with some slight advantages in 
the crossed-beam configuration. This means that, with respect to a 
monochromatic optical dipole trap, a more substantial Bose thermal
cloud can be sustained while the Fermi gas is in a deeper degenerate regime. 
In the various experiments reaching the Fermi degenerate regime, 
the estimate of the temperature is usually obtained by fitting the 
surviving normal Bose component.
As the temperature is lowered, the thermal component shrinks in 
amplitude and size, and the fit to assess its temperature is 
less accurate. 
This effect is mitigated in our bichromatic traps thus allowing 
for a more precise thermometry. 

Simple estimates for the minimum degeneracy parameter
$T^*/T_\mathrm{F}$ can be obtained by extending the qualitative
discussion reported in the concluding part of Section 2.
Below $T_\mathrm{c}$ the heat capacity of an ideal Bose gas can be 
written as \cite{BAGNATO}
\begin{equation}
C_\mathrm{b} \simeq 10.8 ~k_B N_\mathrm{b} \left(\frac{T}{T_\mathrm{c}} \right)^3
\label{CBAPROX}
\end{equation}
while for $T/T_\mathrm{F} \leq 0.5$ with a good approximation the heat 
capacity of a Fermi gas is linearly dependent on the temperature (see
also Figure 9) \cite{BUTTS}
\begin{equation}
C_\mathrm{f} \simeq \pi^2 k_B N_\mathrm{f} \frac{T}{T_\mathrm{F}}.
\label{CFAPROX}
\end{equation}
By taking the ratio of $C_\mathrm{b}$ and $C_\mathrm{f}$ and observing
that according to Eqs. (\ref{TF}) and (\ref{Tc}) $T_\mathrm{c}$ 
can be expressed 
in terms of $T_\mathrm{F}$, we obtain the degeneracy parameter
$T/T_\mathrm{F}$ as
\begin{equation}
\frac{T}{T_\mathrm{F}} \simeq 0.35 
\left( \frac{\omega_\mathrm{b}}{\omega_\mathrm{f}} \right)^{3/2} 
\left( \frac{C_\mathrm{b}}{C_\mathrm{f}} \right)^{1/2}.
\label{TTF}
\end{equation}
In a conservative scenario we can assume that sympathetic cooling
stops when $C_\mathrm{b} \simeq C_\mathrm{f}$. In this case, for 
$\omega_\mathrm{f}/\omega_\mathrm{b} \simeq 1$ (as in the case of the 
$^6$Li-$^7$Li mixture) we get a minimum value 
of the degeneracy parameter at the end of the cooling as
$T^*/T_\mathrm{F}\simeq 0.35$.
If cooling is still possible for $C_\mathrm{b} < C_\mathrm{f}$ more 
optimistic estimates can be given. 
For instance if cooling stops when $C_\mathrm{b}/C_\mathrm{f} \simeq 0.1$, 
then $T^*/T_\mathrm{F} \simeq 0.11$. By using a larger 
$\omega_\mathrm{f}/\omega_\mathrm{b}$ ratio the $T^*/T_\mathrm{F}$ ratio 
decreases according to (20). It is interesting to note that the dependence is 
quite sensitive to the frequency ratio, and even minor deviations of 
this ratio from unity result in an observable effect. For instance, 
in a magnetic trap with a $^6$Li-$^{23}$Na mixture we have 
$\omega_\mathrm{f}/\omega_\mathrm{b} \simeq 
(m_\mathrm{b}/m_\mathrm{f})^{1/2} \simeq 1.96$ and therefore 
we estimate, for $C_\mathrm{b}/C_\mathrm{f}=0.1$, a value 
$T^*/T_\mathrm{F}=0.04$, very close to the minimum value $T/T_\mathrm{F}=0.05$
recently obtained by the MIT group for a complete evaporation 
of the Bose component \cite{HADZIBABIC1}. 

Superfluidity of the Fermi gas is expected below the
critical temperature for the onset of atomic Cooper pairs \cite{GORKOV,STOOF} 
\begin{equation}
T_\mathrm{BCS} = \frac{5}{3e}  e^{-\pi/2k_\mathrm{F}|a_\mathrm{f}|} 
T_\mathrm{F},
\end{equation}
where $k_\mathrm{F}$ is the Fermi wavevector such that 
$E_\mathrm{F}=\hbar^2 k_\mathrm{F}^2/2 m_\mathrm{f}$, and
$a_\mathrm{f}$ is the fermion-fermion elastic scattering length. 
Besides leaving freedom to apply arbitrary homogeneous magnetic fields
to enhance the scattering length through tuning to a Feshbach
resonance \cite{TIMMERMANS1,HOLLAND,OHASHI}, as demonstrated 
experimentally for Fermi gases in \cite{LOFTUS,MODUGNOLAST}, 
our bichromatic configuration allows also for an independent increase 
of $k_\mathrm{F}$ due to the higher achievable densities. 
The resulting $T_\mathrm{BCS}/T_\mathrm{F}$  are within the accessible
range which corresponds, as seen in Figure 8, 
to $T/T_\mathrm{F} \gtrsim 3 \times 10^{-2}$. 
The use of optical trapping also leads to a large absolute value 
of the Fermi temperature which can be otherwise obtained by magnetic 
trapping only with particular geometries maximizing the field 
gradients \cite{SCHRECK0}. Finally, the presence of the Bose gas 
could allow for enhancements of the BCS pairing temperature since 
bosons can mediate phonon-exchange between fermions in a way 
analogous to ordinary superconductors \cite{HEISELBERG,BIJLSMA}. 

In general, with the technique discussed above new mixtures consisting
of a normal Bose gas (or a Bose condensed gas coexisting with a large
Bose thermal fraction) and a degenerate Fermi gas are viable. 
This contrasts the only situation known so far of a ${}^3$He-${}^4$He
mixture where degeneracy is reached earlier for the bosonic species.
One of the advantages of using such an anomalous mixture is the
possibility to have a well controllable background - a normal Bose gas
- superimposed to the Fermi gas. 
This considerably simplifies the possible signatures of a superfluid 
phase transition in a Fermi gas.
For instance it should possible to look at a bulge in the density 
distribution as predicted in \cite{CHIOFALO}, since this is obtained 
admist a smooth, low density and well controllable thermal cloud 
instead of a higher density and peaked condensate \cite{AMORUSO}. 
The presence of a superfluid state could be evidenced also by using 
the same blue-detuned beam as a mechanical stirrer for the fermion 
cloud. In this case one should look at a finite threshold for the 
onset of a dissipative motion or a drag force \cite{RAMAN,ONOFRIO}.
The presence of the bosonic thermal cloud (or a condensate component) 
gives rise to heating for all stirring velocities (or above a critical
velocity depending upon the condensate density) of the laser beam \cite{RAMAN1}.  
However the contribution of the thermal cloud to the heating is very low,
and much smaller than the Rayleigh scattering in the relevant range of 
stirring velocities, due to the low density. 
In order to better discriminate heating coming from the Bose component, 
one could also take advantage of the proposed manipulation of an 
ultracold cloud with Raman beams \cite{HIGBIE,STAMPERKURN} to 
create a directional critical velocity for the Fermi component. 
Another advantage of a Fermi-Bose mixture with the latter component 
in the normal state is the possibility to perform experiments on 
scattering from microscopic impurities 
(such as the one described in \cite{CHIKKATUR} for a Bose
condensate) in a much simpler way than using two isotopes.

The presence of the Bose cooler in the non-condensed phase allows one 
to re-examine also species for which Bose condensation has been proven 
difficult to achieve, like cesium (see however the recent achievement 
of BEC in Cs with purely optical means \cite{GRIMM}). 
For instance, one can reconsider the use of $^{133}$Cs \cite{BOIRON} 
which, due to its large mass and small recoil temperature, can be
efficiently cooled to very low temperature in a magneto-optical trap. 
This would ensure robust initial conditions, in terms
of initial number of atoms and initial temperature, to start an 
efficient evaporative cooling in the optical dipole trap.
Even if sympathetic cooling is less efficient due to the large 
mass ratio with the Fermi species \cite{DELANNOY}, the same feature 
is an advantage in terms of ratio between the trapping frequencies 
even at zero blue-detuned beam intensity or in a purely magnetic trap. 
Recently, sympathetic cooling of $^{7}$Li through $^{133}$Cs has been 
demonstrated in a far-off resonance optical trap \cite{MUDRICH}. 
A CO$_2$ laser was used to create the optical potential obtaining, 
due to the static atomic polarizabilities, a larger energy depth for 
$^{133}$Cs. As a consequence the $^{7}$Li went under simultaneous 
sympathetic and evaporative cooling. 
The problem can be circumvented by using a Nd:YAG laser as the primary, 
red-detuned trapping laser, and a deconfining beam with wavelength 
in between the two atomic transition wavelengths.
Furthermore, by relaxing the requirement for using an alkali species 
(using for instance ytterbium cooled in an optical dipole trap, 
see \cite{HONDA,TAKASU}, and recently brought into the degenerate 
regime for a bosonic isotope \cite{TASAKU1}) various favourable 
possibilities can be envisaged.
 
\section{Conclusions}
A novel path to reach deep Fermi degeneracy through sympathetic
cooling with a Bose gas undergoing evaporative cooling in an optical
dipole trap has been proposed. The key feature is that the trapping
frequencies, determining the degree of degeneracy of the dilute gases,
are made different by properly using a second, deconfining beam. 
Both single- and crossed-beam configurations have been studied for 
the four Fermi-Bose mixtures available with 
stable alkali fermions. In all cases, a substantial increase of the 
Fermi energy is expected maintaining at the same time the critical 
temperature for Bose condensation constant or slightly   
decreased with respect to a single color optical dipole trap.
This decrease of the critical temperature for Bose condensation
is beneficial for maintaining a precision thermometry in 
the deep degenerate regime for the Fermi gas, and it makes also possible 
the use of species which can hardly reach Bose condensation.
With respect to the routes for fermion cooling currently 
under active investigation, our proposal does not necessarily require 
enhancement of elastic scattering lengths through Feshbach 
resonances, a mechanism often associated to the increase of  
inelastic processes rates, in competition with hydrodynamic 
behaviour of the atomic clouds or formation of ultracold molecules
\cite{REGAL,STRECKER,HADZIBABIC1,CUBIZOLLES}, already present even 
in nondegenerate conditions \cite{JOCHIN}. 
With respect to the pure Fermi mixtures, it allows instead for a more 
precise determination of the temperature and, in particular, of the BCS-like phase 
transition temperature below which onset of superfluidity is expected \cite{STOOF}. 
Two technical issues still to be addressed in detail are the 
degree of control of the laser beams to achieve a common focus, 
and the intensity ratio stability between confining and deconfining 
lasers. In a more pessimistic scenario, the proposed technique 
will allow to study the heating mechanisms preventing further cooling 
of fermions, in a way similar to the one proposed in
\cite{TIMMERMANS2}, or non-equilibrium phenomena 
in the ultracold regime analogous to those already explored for the 
Bose condensed gases \cite{SHVARCHUCK}, or proposed as an interesting 
alternative to quasi-equilibrium sympathetic cooling \cite{CARR}.

\section*{Acknowledgments}
We are indebted to our teacher, colleague, and friend Giovanni 
Jona-Lasinio for longstanding, pleasant scientific discussions 
and for having encouraged us in pursuing independent thinking 
at an early stage of our carrier. 
This research was partially supported by Cofinanziamento 
MIUR protocollo 2002027798.
R. O. also acknowledges support by the Department of Energy, 
under Contract No. W-7405-ENG-36. 

\appendix
\section{Ideal Bose and Fermi systems with a finite number of particles
in a harmonic potential}

In this section we review some elementary thermodynamic properties 
of noninteracting quantum systems consisting of a finite number of 
bosonic or fermionic particles trapped into an external potential.
More specifically, we suppose that the particles of mass $m$ are 
confined by the anisotropic harmonic potential
\begin{eqnarray}
V(x,y,z) = \frac{1}{2} m 
\left( \omega_x^2 x^2 + \omega_y^2 y^2 + \omega_z^2 z^2 \right),
\end{eqnarray}\noindent
so that the corresponding single-particle energy levels are given by
\begin{eqnarray}
E_{n_x,n_y,n_z}=
\hbar\omega_x \left( n_x+\frac{1}{2} \right)+
\hbar\omega_y \left(n_y+\frac{1}{2}\right)+
\hbar\omega_z \left(n_z+\frac{1}{2}\right),
\end{eqnarray}
with $n_x,n_y,n_z=0,1,2,\ldots$.

In the grand-canonical ensemble, the average number of particles 
and the energy of the system are, respectively   
\begin{eqnarray}
N &=& 
\sum_{n_x=0}^{\infty}\sum_{n_y=0}^{\infty}\sum_{n_z=0}^{\infty}
\frac{1}
{\exp\left({\frac{E_{n_x,n_y,n_z}-\mu}{k_\mathrm{B} T}}\right) \pm 1}
\label{N}
\\
E &=& 
\sum_{n_x=0}^{\infty}\sum_{n_y=0}^{\infty}\sum_{n_z=0}^{\infty}
\frac{E_{n_x,n_y,n_z}}
{\exp\left({\frac{E_{n_x,n_y,n_z}-\mu}{k_\mathrm{B} T}}\right) \pm 1},
\label{E}
\end{eqnarray}
where $T$ is the temperature of the reservoir, 
$\mu$ the chemical potential, 
and the upper (lower) sign holds for fermions (bosons).
Usually, one knows the temperature $T$ of the reservoir and 
the number $N$ of particles of the system so that the chemical potential 
must be evaluated as a function of $N$ and $T$.
The value of $\mu(N,T)$ is determined by solving the nonlinear 
Eq. (\ref{N}) numerically, e.g. by truncating the series in the r.h.s. 
at a proper order and controlling the convergence error \cite{NAPOLITANO}.
Once $\mu(N,T)$ is known, the energy $E(\mu(N,T),T)$ is evaluated in an
analogous way by truncating the series in the r.h.s. of (\ref{E}).

The computation of the series in (\ref{N}) or (\ref{E}) requires
the evaluation of triply nested sums which may be very time consuming.
A more favourable situation, with double or single series to be computed,
is obtained in the presence of symmetries of the external potential.
For instance, in the case of experimental interest in which 
$\omega_x=\omega_y\equiv\omega_{xy}$, Eq. (\ref{N}) becomes
\begin{eqnarray}
N &=& 
\sum_{n_{xy}=0}^{\infty}\sum_{n_z=0}^{\infty}
\frac{n_{xy}+1}
{\exp\left({\frac{E_{n_{xy},n_z}-\mu}{k_\mathrm{B} T}}\right) \pm 1},
\label{NN}
\end{eqnarray}
where the factor $n_{xy}+1$ represents the degeneracy of the
level at energy 
$E_{n_{xy},n_z}=\hbar\omega_{xy} \left(n_{xy}+1\right)+
\hbar\omega_z \left(n_z+\frac{1}{2}\right)$ 
with $n_{xy},n_z=0,1,2,\ldots$.

The heat capacity of the system at fixed number of particles is 
\begin{eqnarray}
C(N,T) &=& \frac{\mathrm{d}E(N,T)}{\mathrm{d}T} \nonumber \\
&=& \frac{\partial E}{\partial T} + \frac{\partial E}{\partial \mu}
~ \frac{\partial \mu}{\partial T} 
\nonumber \\
&=& \frac{\partial E}{\partial T} - \frac{\partial E}{\partial \mu}
~ \frac{\partial N}{\partial T} 
\left( \frac{\partial N}{\partial \mu} \right)^{-1}, 
\end{eqnarray}
where we used the constraint
$\mathrm{d}N/\mathrm{d}T= \partial N/\partial T + 
(\partial N/\partial \mu) (\partial \mu/\partial T)=0$.
The derivatives $\partial N/\partial T$, $\partial N/\partial \mu$,
$\partial E/\partial T$, and $\partial E/\partial \mu$ are
obtained by (\ref{N}) and (\ref{E}) and can be evaluated
numerically as explained above.  
As an example, in Figure 9 we show the behavior of $C(N,T)$ as a
function of $T$ for different values of the number of particles $N$.
While in the fermion case $C(N,T)$ is almost independent of $N$,
for bosons the size effect is striking around the critical temperature.
\begin{figure}
\begin{center}
\psfrag{T/T_c}{$T/T_\mathrm{c}$}
\psfrag{T/T_F}{$T/T_\mathrm{F}$}
\psfrag{C/Nk_B}{$C/Nk_\mathrm{B}$}
\includegraphics[width=1.0\columnwidth]{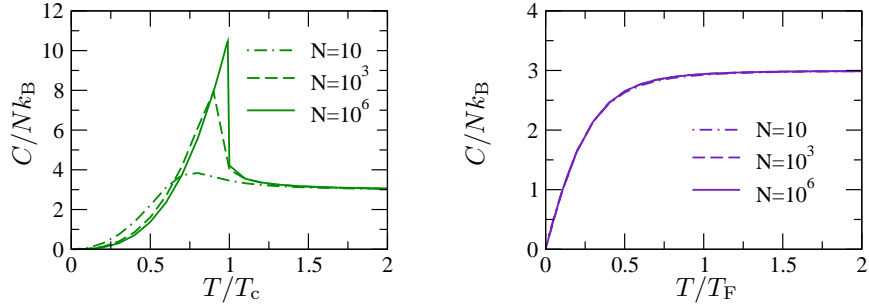}
\caption{
Specific heats of a system of $N$ noninteracting bosons (left panel)
and fermions (right panel) in a harmonic trap as a function of the 
normalized temperatures $T/T_\mathrm{c}$ and $T/T_\mathrm{F}$.
We have chosen $\omega_x=\omega_y=\omega_z/\sqrt{2}$ as 
in a crossed-beam optical dipole trap.}
\end{center}
\end{figure}

Finally, we derive the values of the Fermi and Bose degeneracy
temperatures reported in (\ref{TF}) and (\ref{Tc}).
In the case of fermions, for any value of $N$ and $T$ we have 
$\mu(N,T) < \mu(N,T=0)$ and the Fermi energy is defined as 
$E_\mathrm{F}(N)=\mu(N,T=0)$.
For $k_\mathrm{B}T \gg \hbar \omega$, 
a condition well fulfilled in the experiments, the Fermi 
energy can be evaluated with the approximation
\begin{eqnarray}
N &=& 
\sum_{\genfrac{}{}{0pt}{}{n_x,n_y,n_z}{E_{nx,ny,nz} < E_\mathrm{F}}}
1
\nonumber \\
&\simeq& \frac{\frac{1}{6}E_\mathrm{F}^3}{\hbar \omega_x \hbar \omega_y 
\hbar \omega_z}
= \frac{E_\mathrm{F}^3}{6 \hbar^3 \omega_\mathrm{f}^3},
\end{eqnarray}
where $\omega_\mathrm{f}=\left( \omega_x \omega_y \omega_z \right)^{1/3}$.
Therefore, $k_\mathrm{B}T_\mathrm{F} \equiv E_\mathrm{F} =
(6N)^{1/3} \hbar \omega_\mathrm{f}$.

In the case of bosons, we separate the occupation of the 
ground state from that of the excited ones:
\begin{eqnarray}
N &=& N_0 + N_e 
\nonumber \\ &=& 
\frac{1} {e^{\frac{E_{0}-\mu(N,T)}{k_\mathrm{B} T}} - 1} +
\sum_{\genfrac{}{}{0pt}{}{n_x,n_y,n_z}{\neq 0,0,0}}
\frac{1}
{e^{\frac{E_{n_x,n_y,n_z}-\mu(N,T)}{k_\mathrm{B} T}} - 1},
\end{eqnarray}\noindent
where $\mu(N,T) < E_0 \equiv E_{0,0,0}$. 
When $T \to 0$, we have $\mu \to E_0$ and $N_0 \to N$. 
The critical temperature $T_\mathrm{c}$ is defined by the condition
$N_e(\mu=E_0,T_\mathrm{c}) = N$.
For $k_\mathrm{B}T \gg \hbar \omega$, we have
\begin{eqnarray}
N_e(\mu=E_0,T_\mathrm{c}) &\simeq&
\int_0^\infty \mathrm{d}n_x \mathrm{d}n_y \mathrm{d}n_z~
\frac{1}{e^{\frac{n_x\hbar\omega_x+n_y\hbar\omega_y+n_z\hbar\omega_z}
{k_\mathrm{B}T_\mathrm{c}}}-1}
\nonumber \\ &=& 
\left( \frac{k_\mathrm{B}T_\mathrm{c}}{\hbar \omega_\mathrm{b}} \right)^3
\int_0^\infty \mathrm{d}x \mathrm{d}y \mathrm{d}z~
\frac{1}{e^{x+y+z}-1}
\nonumber \\ &=& 
\left( \frac{k_\mathrm{B}T_\mathrm{c}}{\hbar \omega_\mathrm{b}} \right)^3
\zeta(3),
\end{eqnarray}\noindent
where $\omega_\mathrm{b}=\left( \omega_x \omega_y \omega_z \right)^{1/3}$
and $\zeta$ is the Riemann zeta function.
Therefore, $k_\mathrm{B}T_\mathrm{c} = (N/\zeta(3))^{1/3}
\hbar \omega_\mathrm{b}$.


\begin{thebibliography}{20}

\bibitem{NAMBU} Y. Nambu and G. Jona-Lasinio, Dynamical model of
elementary particles based on an analogy with superconductivity. 
I, Phys. Rev. \textbf{122}: 345 (1961); II, \textbf{124}: 246 (1961).

\bibitem{TILLEY} D. R. Tilley and J. Tilley, {\sl Superfluidity and
Superconductivity}, (Institute of Physics, Bristol and Philadelphia,
1996).

\bibitem{CHU} S. Chu, The manipulation of neutral particles,
Rev. Mod. Phys. \textbf{70}: 685 (1998).

\bibitem{COHEN} C. N. Cohen-Tannouji, Manipulating atoms with
photons, Rev. Mod. Phys. \textbf{70}: 707 (1998).

\bibitem{PHILLIPS} W. D. Phillips, Laser cooling and trapping of
neutral atoms, Rev. Mod. Phys. \textbf{70}: 721 (1998).

\bibitem{KETTERLE} W. Ketterle and N. J. Van Druten, Evaporative cooling of 
trapped atoms, in {\sl Advances in Atomic, Molecular, and Optical
  Physics}, edited by B. Bederson and
H. Walther, Vol. 37 (Academic Press, San Diego, 1996), p. 181.

\bibitem{ANDERSON} M. H. Anderson, J. R. Ensher, M. R. Matthews,
C. E. Wieman, and E. A. Cornell, Observation of Bose-Einstein
condensation in a dilute atomic vapor, Science \textbf{269}: 198 (1995).

\bibitem{DAVIS} K. B. Davis, M.-O. Mewes, M. R. Andrews, N. J. van
Druten, D. S. Durfee, D. M. Kurn, and W. Ketterle, Bose-Einstein
condensation in a gas of sodium atoms, Phys. Rev. Lett. \textbf{75}: 3969
(1995).

\bibitem{BRADLEY} C. C. Bradley, C. A. Sackett, J. J. Tollett, and
R. G. Hulet, Evidence of Bose-Einstein condensation in an atomic gas
with attractive interactions, Phys. Rev. Lett. \textbf{75}: 1687
(1995); {\it ibidem} \textbf{79}: 1170 (1997).

\bibitem{ASPECT} {\sl Bose-Einstein condensates and atom lasers},
C. R. Acad. Sci. Paris, t. 2, S\`erie IV, no. 3 (2001), Special issue
edited by A. Aspect and J. Dalibard.

\bibitem{PETHICK} C. J. Pethick and H. Smith, {\sl Bose-Einstein
condensation in dilute gases}, (Cambridge University Press,
Cambridge, 2002).

\bibitem{PITAEVSKIIBOOK} L. P. Pitaevskii and S. Stringari, 
{\sl Bose-Einstein condensation}, (Oxford Science Publications, Oxford, 2003).

\bibitem{MATTHEWS} M. R. Matthews, B. P. Anderson, P. C. Kaljan, 
D. S. Hall, C. E. Wieman, and E. A. Cornell, Vortices in a
Bose-Einstein condensate, Phys. Rev. Lett. \textbf{83}: 2498 (1999).

\bibitem{MADISON} K. W. Madison, F. Chevy, W. Wohlleben, and
J. Dalibard, Vortex formation in a stirred Bose-Einstein condensate,
Phys. Rev. Lett. \textbf{84}: 806 (2000).

\bibitem{ABOSHAEER} J. R. Abo-Shaeer, C. Raman, J. M. Vogels, and
W. Ketterle, Observation of vortex lattices in Bose-Einstein
condensates, Science \textbf{292}: 476 (2001).

\bibitem{MARAGO} O. M. Marag\`o, S. A. Hopkins, J. Arlt, E. Hodby, 
G. Hechenblaikner, and C. J. Foot, Observation of the
scissor mode and evidence for superfluidity of a trapped
Bose-Einstein condensed gas, Phys. Rev. Lett. \textbf{84}: 2059 (2000).

\bibitem{RAMAN} C. Raman, M. K\"ohl, R. Onofrio, D. S. Durfee,
C. E. Kuklewicz, Z. Hadzibabic, and W. Ketterle, Evidence for a
critical velocity in a Bose-Einstein condensed gas,
Phys. Rev. Lett. \textbf{84}: 2502 (1999).

\bibitem{ONOFRIO} R. Onofrio, C. Raman, J. M. Vogels, J. Abo-Shaeer,
A. P. Chikkatur, and W. Ketterle, Observation of a superfluid flow in
a Bose-Einstein condensed gas, Phys. Rev. Lett. \textbf{85}: 2228 (2000).

\bibitem{FETTER} A. L. Fetter and A. A. Svidzinsky, Vortices in a 
trapped dilute Bose-Einstein condensate, J. Phys. Condens. Matter 
\textbf{13}: R135 (2001).

\bibitem{LEGGETT} A. J. Leggett, Bose-Einstein condensation in the 
alkali gases: Some fundamental concepts, Rev. Mod. Phys. \textbf{73}: 
307 (2001).

\bibitem{PITAEVSKIISCIENCE} L. P. Pitaevskii and S. Stringari, 
The quest for superfluidity in Fermi gases, Science \textbf{298}: 2144
(2001).  

\bibitem{GORKOV} L. P. Gor'kov and T. K. Melik-Barkhudarov,
  Contribution to the theory of superfluidity in an imperfect Fermi
gas, Zh. Eksp. Teor. Fiz \textbf{40}: 1452 (1961) [Sov. Phys. JETP 
\textbf{13}: 1018 (1961)]. 

\bibitem{STOOF} H. T. C. Stoof, M. Houbiers, C. A. Sackett, and
R. G. Hulet, Superfluidity of spin-polarized Li-6,
Phys. Rev. Lett. \textbf{76}: 10 (1996).

\bibitem{ONPRE} R. Onofrio and C. Presilla, Reaching Fermi degeneracy
in two-species optical dipole traps, Phys. Rev. Lett. \textbf{89}: 100401
(2002).

\bibitem{DEMARCO} B. DeMarco and D. S. Jin, Onset of Fermi degeneracy
in a trapped atomic gas, Science \textbf{285}: 1703 (1999).

\bibitem{TIMMERMANS} E. Timmermans and R. Cot\`e, Superfluidity in
sympathetic cooling with atomic Bose-Einstein condensates,
Phys. Rev. Lett. \textbf{80}: 3419 (1998).

\bibitem{TRUSCOTT} A. G. Truscott, K. E Strecker, W. I. McAlexander,
G. B. Partridge, and R. G. Hulet, Observation of Fermi pressure in a
gas of trapped atoms, Science \textbf{291}: 2570 (2001).

\bibitem{SCHRECK} F. Schreck, L. Khaykovich, K. L. Corwin, G. Ferrari,
T. Bourdel, J. Cubizolles, and C. Salomon, Quasipure Bose-Einstein
condensate immersed in a Fermi sea, Phys. Rev. Lett. \textbf{87}: 080403
(2001).

\bibitem{GRANADE} S. R. Granade, M. E. Gehm, K. M. O'Hara, and
J. E. Thomas, All-optical production of a degenerate Fermi gas,
Phys. Rev. Lett. \textbf{88}: 120405 (2002).

\bibitem{HADZIBABIC} Z. Hadzibabic, C. A. Stan, K. Dieckmann, S. Gupta,
M. W. Zwierlein, A. G\"orlitz, and W. Ketterle, Two-species mixture
of quantum degenerate Bose and Fermi gases, Phys. Rev. Lett. 
\textbf{88}: 160401 (2002).

\bibitem{ROATI} G. Roati, F. Riboli, G. Modugno, and M. Inguscio,
Fermi-Bose quantum degenerate $^{40}$K-$^{87}$Rb mixture with 
attractive interaction, Phys. Rev. Lett. \textbf{89}: 150403 (2002).

\bibitem{GEHM} M. E. Gehm, S. L. Hemmer, S. R. Granade, K. M. O'Hara,
  and J. E. Thomas, Mechanical stability of a strongly interacting
  Fermi gas of atoms, Phys. Rev. A \textbf{68}: 011401(R) (2003).

\bibitem{REGAL} C. A. Regal, C. Tickner, J. L. Bohn, and D. S. Jin, 
Creation of ultracold molecules from a Fermi gas of atoms, 
Nature \textbf{424}: 47 (2003).

\bibitem{STRECKER} K. E. Strecker, G. B. Partridge, and R. G. Hulet, 
Conversion of an atomic Fermi gas to a long-lived molecular Bose gas, 
Phys. Rev. Lett. \textbf{91}: 080406 (2003). 

\bibitem{HADZIBABIC1} Z. Hadzibabic, S. Gupta, C. A. Stan,
  C. H. Schunck, M. W. Zwierlein, K. Dieckmann, and W. Ketterle, 
Phys. Rev. Lett. \textbf{91}: 160401 (2003). 

\bibitem{CUBIZOLLES} J. Cubizolles, T. Bourdel, S.J.J.M.F. Kokkelmans, 
G. V. Shlyapnikov, and C. Salomon, 
Phys. Rev. Lett. \textbf{91}: 240401 (2003). 

\bibitem{GOLDWIN} J. Goldwin, S. B. Papp, B. DeMarco, and D. S. Jin,
Two-species magneto-optical trap with $^{40}$K and $^{87}$Rb,
Phys. Rev. A \textbf{65}: 021402 (2001).

\bibitem{FESHBACH} H. Feshbach, A unified theory of nuclear reactions
II, Ann. Phys. (NY) \textbf{19}: 287 (1962).

\bibitem{TIMMERMANSREV} E. Timmermans, P. Tommasini, M. Hussein, 
and A. Kerman, Feshbach resonances in atomic Bose-Einstein condensates,
Phys. Rep. \textbf{315}: 199 (1999).

\bibitem{INOUYE} S. Inouye, M. R. Andrews, J Stenger, H.J. Miesner, 
D. M. Stamper-Kurn, and W. Ketterle, Observation of Feshbach resonances in 
a Bose-Einstein condensate, Nature \textbf{392}: 151 (1998).

\bibitem{COURTEILLE} Ph. Courteille, R. S. Freeland, D.J. Heinzen, 
F. A. van Abeelen, B. J. Verhaar, Observation of a Feshbach resonance 
in cold atom scattering, Phys. Rev. Lett. \textbf{81}: 69 (1998).

\bibitem{ROBERTS} J. L. Roberts, N. R. Claussen, J. P. Burke, Jr., 
C. H. Greene, E. A. Cornell, and C. E. Wieman, Resonant magnetic field 
control of elastic scattering in cold $^{85}$Rb, Phys. Rev. Lett. 
\textbf{81}: 5109 (1998).

\bibitem{VULETIC} V. Vuletic, A. J. Kerman, C. Chin, S. Chu, 
Observation of low-field Feshbach resonances in collision of Cesium 
atoms, Phys. Rev. Lett. \textbf{82}: 1406 (1999).

\bibitem{TIMMERMANS1} E. Timmermans, V. Furuya, P. W. Milonni, and
A. K. Kerman, Prospect of creating a composite Fermi-Bose superfluid,
Phys. Lett. A \textbf{285}: 228 (2001).

\bibitem{HOLLAND} M. Holland, S.J.J.M.F. Kokkelmans, M. L. Chiofalo,
and R. Walser, Resonance superfluidity in a quantum degenerate gas,
Phys. Rev. Lett. \textbf{87}: 120406 (2001).

\bibitem{CHIOFALO} M. L. Chiofalo, S.J.J.M.F. Kokkelmans,
J. N. Milstein, and M. J. Holland, Signatures of resonance
superfluidity in a quantum Fermi gas, Phys. Rev. Lett. \textbf{88}:
120406 (2002).

\bibitem{OHASHI} Y. Ohashi and A. Griffin, 
BCS-BEC crossover in a gas of Fermi atoms with a Feshbach resonance,
Phys. Rev. Lett. \textbf{89}: 130402 (2002).

\bibitem{ASKIN} A. Askin and J. Gordon, 
Cooling and trapping of atoms by resonance radiation pressure,
Opt. Lett. \textbf{4}: 161 (1979).

\bibitem{GORDON} J. P. Gordon and A. Askin, Motion of atoms in a
radiation trap, Phys. Rev. A \textbf{21}: 1606 (1980).

\bibitem{CHU1} S. Chu, J. E. Bjorkholm, A. Askin, and A. Cable,
Experimental observation of optically trapped atoms,
Phys. Rev. Lett. \textbf{57}: 314 (1986).

\bibitem{MILLER} J. Miller, R. Cline, and D. Heinzen, Far-off resonance 
optical trapping of atoms, Phys. Rev. A \textbf{47}: R4567 (1993).

\bibitem{ADAMS} C. S. Adams, H. J. Lee, N. Davidson, M. Kasevich, 
and S. Chu, Evaporative cooling in a crossed dipole trap,
Phys. Rev. Lett. \textbf{74}: 3577 (1995).

\bibitem{STAMPER} D. M. Stamper-Kurn, M. R. Andrews, A. P. Chikkatur,
S. Inouye, H.-J. Miesner, J. Stenger, and W. Ketterle, Optical
confinement of a Bose-Einstein condensate, Phys. Rev. Lett. \textbf{80}: 2027 (1998).

\bibitem{BARRETT} M. D. Barrett, J. A. Sauer, and M. S. Chapman,
All-optical formation of an atomic Bose-Einstein condensate,
Phys. Rev. Lett. \textbf{87}: 010404 (2001).

\bibitem{GRIMM} R. Grimm, M. Weidem\"uller, and Y. B. Ovchinkov, 
Optical dipole traps for neutral atoms, Ad. At. Mol. Op. \textbf{42}: 95 (2000).

\bibitem{LOFTUS} T. Loftus, C. A. Regal, C. Tickner, J. L. Bohn, and
D. S. Jin, Resonant control of elastic collisions in an optically
trapped Fermi gas of atoms, Phys. Rev. Lett. \textbf{88}: 173201 (2002).

\bibitem{MODUGNOLAST} G. Modugno, G. Roati, F. Riboli, F. Ferlaino, 
R. J. Brecha, and M. Inguscio, Collapse of a degenerate Fermi gas,
Science \textbf{297}: 2240 (2002).

\bibitem{OHARA0} K. M. O'Hara, S. L. Hemmer, M. E. Gehm,
S. R. Granade, J. E. Thomas, Observation of a strongly interacting 
degenerate Fermi gas of atoms, Science \textbf{298}: 2179 (2002).

\bibitem{NOTETTF} In this experiment the lowest $T/T_\mathrm{F}$ ratio with 
dual evaporative cooling of two hyperfine fermion states has been achieved, 
as $0.08 < T/T_\mathrm{F} < 0.18$. 
By performing an identical evaporation process on an equally populated 
mixture of the two hyperfine states the heat capacities are always well matched. 
On the other hand any determination of the temperature for a system made of 
degenerate fermions is strongly model dependent and drops in sensitivity at 
the lowest explored temperatures, as witnessed by the relatively 
large range of values for the estimated $T/T_\mathrm{F}$ ratio.

\bibitem{MENOTTISTRING} C. Menotti, P. Pedri, and S. Stringari, 
Expansion of an interacting Fermi gas, Phys. Rev. Lett. \textbf{89}:
250402 (2002).

\bibitem{PRESILLA} C. Presilla and R. Onofrio, 
Cooling dynamics of ultracold two-species Fermi-Bose mixtures, 
Phys. Rev. Lett. \textbf{90}: 030404 (2003).

\bibitem{VIVERIT} L. Viverit, S. Giorgini, L. P. Pitaevskii, and
S. Stringari, Adiabatic compression of a trapped Fermi gas,
Phys. Rev. A \textbf{63}: 033603 (2001).

\bibitem{THOMAS} N. R. Thomas, A. C. Wilson, C. J. Foot, 
Double-well magnetic trap for Bose-Einstein condensation,
Phys. Rev. A \textbf{65}: 063406 (2002).

\bibitem{SCHUENEMANN} U. Sch\"unemann, H. Engler, M. Zielonkowski,
M. Weidem\"uller, and R. Grimm, Magneto-optic trapping of lithium
using semiconductor lasers, Optics Communications \textbf{158}: 263
(1998).

\bibitem{MODUGNO1} G. Modugno, C. Benk\"o, P. Hannaford, 
G. Roati, and M. Inguscio, Sub-Doppler laser cooling of fermionic 
${}^{40}$K atoms, Phys. Rev. A \textbf{60}: R3373 (1999).

\bibitem{TAKEKOSHI1} T. Takekoshi, J. R. Yeh, R. J. Knize,
Quasi-electrostatic trap for neutral atoms, Optics Communications {\bf
114}: 421 (1995).

\bibitem{TAKEKOSHI2} T. Takekoshi and R. J. Knize, CO$_2$ laser trap
for cesium atoms, Optics Letters \textbf{21}: 77 (1996).

\bibitem{OHARA} K. M. O'Hara, S. R. Granade, M. E. Gehm, T. A. Savard,
S. Bali, C. Freed, and J. E. Thomas, Ultrastable CO$_2$ laser
trapping of lithium atoms, Phys. Rev. Lett. \textbf{82}: 4204 (1999).

\bibitem{OHARA1} K. M. O'Hara, S. R. Granade, M. E. Gehm, and
J. E. Thomas, Loading dynamics of CO$_2$ laser traps, Phys. Rev. A
\textbf{63}: 043403 (2001).

\bibitem{SAVARD} T.A. Savard, K.M. O'Hara, and J. E. Thomas, 
Laser-noise-induced heating in far-off resonance optical traps, 
Phys. Rev. A \textbf{56}: R1095 (1997). 

\bibitem{OHARA2} K. M. O'Hara, M. E. Gehm, S.R. Granade, and
J. E. Thomas, Scaling laws for evaporative cooling in time-dependent
optical traps, Phys. Rev. A \textbf{64}: 051403R (2001).

\bibitem{HOUBIERS} M. Houbiers, H. T. C. Stoof, W. I. McAlexander,
and R. G. Hulet, Elastic and inelastic collisions of Li-6 atoms in
magnetic and optical traps, Phys. Rev. A \textbf{57}: R1497 (1998).

\bibitem{NOTE1} In the case of magnetic trapping for alkali species
confinement and evaporation are instead kept separated. It is then 
possible to further compress the atomic cloud during rf-induced
evaporation, which explains the earlier success of this technique to
achieve the Bose-degenerate regime.

\bibitem{LARSON} D. J. Larson, J. C. Bergquist, J. J. Bollinger, 
W. M. Itano, and D. J. Wineland, Sympathetic cooling of trapped ions:
A laser-cooled two-species nonneutral ion plasma, 
Phys. Rev. Lett. \textbf{57}: 70 (1986).

\bibitem{DECARVALHO} R. deCarvalho, J. M. Doyle, B. Friedrich,
T. Guillet, J. Kim, D. Patterson, and J. D. Weinstein, 
Buffer-gas loaded magnetic traps for atoms and molecules: A primer, 
Eur. Phys. J. D \textbf{7}: 289 (1999).

\bibitem{MYATT} C. Myatt, E. Burt, R. Ghrist, E. Cornell, and
C. Wieman, Production of two overlapping Bose-Einstein condensates 
by sympathetic cooling, Phys. Rev. Lett. \textbf{78}: 586 (1997). 

\bibitem{BLOCH} I. Bloch, M. Greiner, O. Mandel, T. W. H\"ansch, 
and T. Esslinger, Sympathetic cooling of $^{85}$Rb and $^{87}$Rb, 
Phys. Rev. A \textbf{64}: 021402(R) (2001).

\bibitem{MODUGNO} G. Modugno, G. Ferrari, G. Roati, A. Simoni, and 
M. Inguscio, Bose-Einstein condensation of potassium atoms by 
sympathetic cooling, Science \textbf{294}: 1320 (2001).

\bibitem{CHIKKATUR} A. Chikkatur, A. G\"orlitz, D. M. Stamper-Kurn, 
S. Inouye, S. Gupta, and W. Ketterle, Suppression and enhancement
of impurity scattering in a Bose-Einstein condensate,
Phys. Rev. Lett. \textbf{85}: 483 (2000).

\bibitem{GIORGINI} S. Giorgini, L. P. Pitaevskii, and S. Stringari, 
Thermodynamics of a trapped Bose-condensed gas, 
J. Low Temp. Phys. \textbf{109}: 309 (1997).

\bibitem{BAGNATO} V. Bagnato, D. E. Pritchard, and D. Kleppner, 
Bose-Einstein condensation in an external potential, 
Phys. Rev. A \textbf{35}: 4354 (1987). 

\bibitem{BUTTS} D. A. Butts and D. S. Rokhsar, Trapped Fermi gases, 
Phys. Rev. A \textbf{55}: 4346 (1997).

\bibitem{SCHRECK0} F. Schreck, G. Ferrari, K. L. Corwin, 
J. Cubizolles, L. Khaykovich, M.-O. Mewes, and C. Salomon, 
Sympathetic cooling of bosonic and fermionic lithium gases 
towards quantum degeneracy, Phys. Rev. A \textbf{64}: 011402(R)
(2001).

\bibitem{HEISELBERG} H. Heiselberg, C. J. Pethick, H. Smith, and 
L. Viverit, Influence of induced interaction on the superfluid 
transition in dilute Fermi gases, Phys. Rev. Lett. \textbf{85}: 2418 (2000).

\bibitem{BIJLSMA} M. J. Bijlsma, B. A. Heringa, and H.T.C. Stoof, 
Phonon exchange in dilute Fermi-Bose mixtures: Tailoring the 
Fermi-Fermi interaction, Phys. Rev. A \textbf{61}: 053601 (2000). 

\bibitem{AMORUSO} M. Amoruso, A. Minguzzi, S. Stringari, M. P. Tosi,
and L. Vichi, Temperature-dependent density profiles of trapped
boson-fermion mixtures, Eur. Phys. J. D \textbf{4}: 261 (1998).

\bibitem{RAMAN1} C. Raman, R. Onofrio, J. M. Vogels, J. R. Abo-Shaeer,
and W. Ketterle, Dissipationless flow and superfluidity in gaseous
Bose-Einstein condensates, J. Low Temp. Phys. \textbf{122}: 99 (2001).

\bibitem{HIGBIE} 
J. Higbie and D.M. Stamper-Kurn, Periodically dressed Bose-Einstein 
condensate: A superfluid with an anisotropic and variable critical 
velocity, Phys. Rev. Lett. \textbf{88}: 090401 (2002).

\bibitem{STAMPERKURN} D. M. Stamper-Kurn, Anisotropic dissipation 
of superfluid flow in a periodically dressed Boes-Einstein condensate,
New J. Phys. \textbf{5}: 50 (2003).

\bibitem{BOIRON} D. Boiron, A. Michaud, J. M. Fournier, L. Simard, 
M. Sprenger, G. Grynberg, and C. Salomon, 
Cold and dense cesium clouds in far-detuned dipole traps, 
Phys. Rev. A \textbf{57}: R4106 (1998).

\bibitem{DELANNOY} G. Delannoy, S. G. Murdoch, V. Boyer, 
V. Josse, P. Bouyer, and A. Aspect, Understanding the
production of dual Bose-Einstein condensation with sympathetic
cooling, Phys. Rev. A \textbf{63}: 051602 (2001).

\bibitem{MUDRICH} M. Mudrich, S. Kraft, K. Singer, R. Grimm, A. Mosk, 
and M. Weidem\"uller, Sympathetic cooling with two atomic species 
in an optical trap, Phys. Rev. Lett. \textbf{88}: 253001 (2002). 

\bibitem{HONDA} K. Honda, Y. Tasaku, T. Kuwamoto, M. Kumakura, 
Y. Takahashi, and T. Yabuzaki, Optical dipole force trapping of a 
fermion-boson mixture of ytterbium isotopes, Phys. Rev. A \textbf{66}: 
021401(R) (2002). 

\bibitem{TAKASU} Y. Takasu, K. Honda, K. Komori, T. Kuwamoto,
  M. Kumakura, Y. Takahashi, and T. Yabuzaki, 
High-density trapping of cold ytterbium atoms by an optical dipole
  force, Phys. Rev. Lett. \textbf{90}: 023003 (2003).

\bibitem{TASAKU1} Y. Tasaku, K. Maki, K. Komori, T. Tamako, K. Honda, 
M. Kumakura, T. Yabuzaki, and Y. Takahashi, Spin-singlet Bose-Einstein 
condensation of two-electron atoms, Phys. Rev. Lett. \textbf{91}:
040404 (2003).

\bibitem{JOCHIN} S. Jochim, M. Bartestein, A. Altmeyer, G. Hendl, 
C. Chin, J. Hecker Denschlag, and R. Grimm, 
 Phys. Rev. Lett. \textbf{91}: 240402 (2003).

\bibitem{TIMMERMANS2} E. Timmermans, Degenerate Fermion gas heating 
by hole creation, Phys. Rev. Lett. \textbf{87}: 240403 (2001). 
This interesting mechanism is based on scattering between atoms in the 
Fermi sea and the energetic background atoms due to the residual
pressure in the trapping region, as a sort of Auger effect on the 
trapped Fermi gas. The requirements for the residual background 
pressure ($<10^{-11}$ Torr) to make this heating source negligible are already 
well fulfilled by several ultracold atom experimental apparatuses.
However, the mechanism can severely limit sympathetic cooling of Fermi atoms 
due to the collisions experienced with the Bose atoms. 

\bibitem{SHVARCHUCK} I. Shvarchuck, Ch. Buggle, D. S. Petrov, 
K. Dieckmann, M. Zielonkowski, M. Kemmann, T. Tiecke, W. von Klitzing, 
G. V. Shlyapnikov, and J. T. M. Walraven,  Bose-Einstein condensation into 
non-equilibrium states studied by condensate focusing, 
Phys. Rev. Lett. \textbf{89}: 270404 (2002). 

\bibitem{CARR} L. D. Carr, T. Bourdel, and Y. Castin, Limits of 
sympathetic cooling of fermions by bosons due to particle losses, 
preprint arXiv:cond-mat/0305441 (19 May 2003).

\bibitem{NAPOLITANO} R. Napolitano, J. De Luca, V. S. Bagnato, 
and G. C. Marques, Effect of a finite number of particles in the 
Bose-Einstein condensation of a trapped gas,
Phys. Rev. A \textbf{55}: 3954 (1997).

\end{thebibliography}
\end{document}